\documentclass[journal]{IEEEtran}
\usepackage{amssymb}
\usepackage{amsmath}
\usepackage{bbm}
\usepackage{array}
\usepackage{algorithm}
\usepackage{algorithmic}
\usepackage[super]{nth}
\usepackage{graphicx}
\usepackage{enumerate}
\usepackage{url}
\usepackage{cite}
\usepackage{subcaption}
\usepackage{placeins}

\newcommand{\cS}{\mathcal{S}}
\newcommand{\cA}{\mathcal{A}}

\newcommand{\cN}{\mathcal{N}}
\newcommand{\cE}{\mathcal{E}}
\newcommand{\cC}{\mathcal{C}}

\newcommand{\rf}{\mathrm{f}}
\newcommand{\rN}{\mathrm{N}}
\newcommand{\rw}{\mathrm{w}}

\newcommand{\be}{\begin{equation}}
\newcommand{\ee}{\end{equation}}

\newcommand{\Exp}{\mathbb{E}}

\newcommand{\M}{\mathbb{M}}

\newcommand{\ind}[1]{\mathbbm{1}_{\{#1\}}}   

\begin{document}

\title{Deep Reinforcement Learning for Adaptive Network Slicing in 5G for Intelligent Vehicular Systems and Smart Cities}
\author{\IEEEauthorblockN{Almuthanna Nassar, and Yasin Yilmaz,~\IEEEmembership{Senior Member,~IEEE}}\\
\IEEEauthorblockA{Electrical Engineering Department, University of South Florida, Tampa, FL 33620, USA}
\\E-mails: \{atnassar@usf.edu; yasiny@usf.edu\}
}
\maketitle


\begin{abstract}
Intelligent vehicular systems and smart city applications are the fastest growing Internet of things (IoT) implementations at a compound annual growth rate of 30\%. In view of the recent advances in IoT devices and the emerging new breed of IoT applications driven by artificial intelligence (AI), fog radio access network (F-RAN) has been recently introduced for the fifth generation (5G) wireless communications to overcome the latency limitations of cloud-RAN (C-RAN). We consider the network slicing problem of allocating the limited resources at the network edge (fog nodes) to vehicular and smart city users with heterogeneous latency and computing demands in dynamic environments. We develop a network slicing model based on a cluster of fog nodes (FNs) coordinated with an edge controller (EC) to efficiently utilize the limited resources at the network edge. For each service request in a cluster, the EC decides which FN to execute the task, i.e., locally serve the request at the edge, or to reject the task and refer it to the cloud. We formulate the problem as infinite-horizon Markov decision process (MDP) and propose a deep reinforcement learning (DRL) solution to adaptively learn the optimal slicing policy. The performance of the proposed DRL-based slicing method is evaluated by comparing it with other slicing approaches in dynamic environments and for different scenarios of design objectives. Comprehensive simulation results corroborate that the proposed DRL-based EC quickly learns the optimal policy through interaction with the environment, which enables adaptive and automated network slicing for efficient resource allocation in dynamic vehicular and smart city environments.
\end{abstract}


\section{Introduction}
\label{introduction}

The fifth generation (5G) wireless communication systems will enable massive Internet of Things (IoT) with deeper coverage, very high data rates of multi giga-bit-per-second (Gbps), ultra-low latency, and extremely reliable mobile connectivity \cite{r2,r3}. It is anticipated that the IoT devices will constitute the 50\% of the 29.3 billion connected devices globally by 2023, where  Internet of Vehicles (IoV) and smart city applications are the fastest growing IoT implementations at annual growth rates of $30\%$ and $26\%$, respectively \cite{cisco}. The emerging new breed of IoT applications which involve video analytics, augmented reality (AR), virtual reality (VR), and artificial intelligence (AI) will produce an annual worldwide data volume of 4.8 zettabyte by 2022, which is more than 180 times the data traffic in 2005 \cite{access}. Equipped with variety of sensors, radars, lidars, ultra-high definition (UHD) video cameras, GPS, navigation system, and infotainment facilities, a connected and autonomous vehicle (CAV) will generate $4.0$ terabyte of data in a single day, of which $1.0$ gigabyte need to be processed every second \cite{intel}.
\subsection{Cloud and Fog RAN}
Through centralization of network functionalities via virtualization, cloud radio access network (C-RAN) architecture is proposed to address the big data challenges of massive IoT. In C-RAN, densely-deployed disseminated remote radio units (RRUs) are connected through high capacity fronthaul trunks to a powerful cloud controller (CC) where they share a vast pooling of storage and baseband units (BBUs) \cite{cran}. The centralized computing, processing, and collaborative radio in C-RAN improves network security, flexibility, availability, and spectral efficiency. It also simplifies network operations and management, enhances capacity, and reduces energy usage \cite{energy}. However, considering the fast growing demands of IoT deployments, C-RAN lays overwhelming onus on cloud computing and fronthaul links, and dictates unacceptable delay caused mainly by the large return transmission times, finite-capacity fronthaul trunks, and flooded cloud processors \cite{limitedfh}.
The latency limitation in C-RAN makes it challenging to meet the desired quality-of-service (QoS) requirements, especially for the delay-sensitive IoV and smart city applications \cite{delay}. Hence, an evolved architecture, fog RAN (F-RAN) is introduced to extend the inherent operations and services of cloud to the edge \cite{fogsurvey}. In F-RAN, the fog nodes (FNs) are not only restricted to perform the regular radio frequency (RF) functionalities of RRUs, but they are also equipped with computing, storage, and processing resources to afford the low latency demand by delivering network functionalities directly at the edge and independently from the cloud \cite{foglatency}. However, due to their limited resources compared to the cloud, FNs are unable to serve all requests from IoV and smart city applications, and hence they should utilize their limited resources intelligently to satisfy the QoS requirements in synergy and complementarity with the cloud \cite{joint}.
\subsection{Network Slicing for Heterogeneous IoV and Smart City Demands}
IoV and smart city applications demand various computing, throughput, latency, availability, and reliability requirements to satisfy a desired level of QoS. For instance, in-vehicle audio, news, and video infotainment services are satisfied by the traditional mobile broadband (MBB) services of high throughput and capacity with latency greater than $100$ ms \cite{tactile}. Cloud computing plays an essential role for such delay-tolerant applications. Other examples of delay-tolerant applications include smart parking \cite{smartparking}, intelligent waste management \cite{smartwaste}, infrastructure (e.g., bridges, railways, etc.) monitoring \cite{smartstructure}, air quality management \cite{smartair}, noise monitoring \cite{smartnoise}, smart city lighting \cite{smartlighting}, smart management of city energy consumption \cite{smartenergy}, and automation of public buildings such as schools, museums, and administration offices to automatically and remotely control lighting and air condition \cite{smartpublic}. 

On the other hand, latency and reliability are more critical for other IoV and smart city applications. For instance, deployment scenarios based on enhanced mobile broadband (eMBB) require latency of $4.0$ ms. Enhanced vehicle-to-everything (eV2X) applications demand $3$-$10$ ms latency with packet loss rate of $10^{-5}$. Ultra-reliable and low-latency communications (URLLC) seek latency level of $0.5$-$1.0$ ms and $99.999\%$ reliability \cite{3gpp,5greq}, e.g., autonomous driving \cite{autonomous}. AI-driven and video analytics services are considered both latency-critical and compute-intensive applications \cite{r16}. For instance, real-time video streaming for traffic management in intelligent transportation system (ITS) \cite{smarttraffic} requires a frame rate of $100$ Hz, which corresponds to a latency of $10$ ms between frames \cite{tactile}. Future electric vehicles (EVs) and CAVs are viewed as computers on wheels (COWs) rather than cars because they are equipped with super computers to execute extremely intensive computing tasks including video analytics and AI-driven functionalities. However, with the high power consumption associated with such intense computing, COWs capabilities are still bounded in terms of computing power, storage, and battery life. Hence, computing offloading to fog and cloud networks is inevitable \cite{compoffload}. Especially in a dynamic traffic and load profiles of dense IoV and smart city service requests with heterogeneous latency and computing needs, partitioning RAN resources virtually, i.e., network slicing \cite{101}, assures service customization.

Network slicing is introduced for the evolving 5G and beyond communication technologies as a cost-effective solution for mobile operators and service providers to satisfy various user QoS \cite{globalsip}. In network slicing, a heterogeneous network of various access technologies and QoS demands that share a common physical infrastructure is logically divided into virtual network slices to improve network flexibility. Each network slice acts as an independent end-to-end network and supports various service requirements and a variety of business cases and applications. In this work, we consider the network slicing problem of adaptively allocating the limited edge computing and processing resources in F-RAN to dynamic IoV and smart city applications with heterogeneous latency demands and diverse computing loads.
\subsection{Contributions}
Motivated by satisfying the QoS requirements of the URLLC applications in F-RAN for intelligent vehicular systems and smart city, we provide a novel network slicing framework for sequentially allocating the FNs' limited resources at the network edge to various vehicular and smart city applications with heterogeneous latency needs in dynamic traffic and load profiles while efficiently utilizing the edge resources. Specifically, our contributions can be listed as:
\begin{itemize}
\item Developing a network slicing model based on \textit{edge clustering} to efficiently utilize the computing and signal processing resources at the edge, and proposing a Markov decision process (MDP) formulation for the considered network slicing problem.
\item Providing theoretical basis for learning the optimal MDP policies using deep reinforcement learning (DRL) methods, and showing how to implement deep Q-networks (DQN), a popular DRL method, to adaptively learn the optimal network slicing policy which ensures both user QoS demands and efficient resource utilization.
\item Presenting extensive simulation results to examine the performance and adaptivity of the proposed DQN-based network slicing method in diverse intelligent vehicular systems and smart city environments.
\end{itemize}
\section{Related Work}
\label{literature}
There is an increasing number of works in the literature focusing on network slicing as an emerging network architecture for 5G and future technologies. Issues and challenges of network slicing as well as the key techniques and solutions for resource management are considered in \cite{101}. The work in \cite{102} provides an overview of various use cases and network requirements of network slicing. Network slicing for resource allocation in F-RAN is considered in \cite{105,106,111}, where network is logically partitioned into two slices, a high downlink-transmission-rate slice for MBB applications, and a low-latency slice to support URLLC services. While \cite{105} focuses on efficiently allocating radio resources and satisfying various QoS requirements, \cite{106} investigates a joint radio and caching resource allocation problem. For massive IoT environment, the authors in \cite{111} proposed a hierarchical architecture in which a global resource manager allocates the radio resources to local resource managers in slices, which assign resources afterwards to their users. Two-level resource management in C-RAN is explored in \cite{113,114}: an upper level for allocating fronthaul capacity and computing resources of C-RAN among multiple wireless operators, and a lower level for controlling the allocation of C-RAN radio resources to individual operators.

Reinforcement learning (RL) is embraced as a powerful tool to deal with dynamic network slicing for adaptive resource allocation in F-RAN. In \cite{icc,globalsip,access}, the RL methods of Q-learning (QL), Monte Carlo, SARSA, expected SARSA, and dynamic programming are utilized to learn the optimal resource allocation policy for a single fog node. The work \cite{incumbent} follows the problem formulation in \cite{access} with an extension to spectrum sharing between 5G users and incumbent users. As the complexity of the control problem increases with more fog nodes, deep RL (DRL), which integrates deep neural networks (DNN) with RL, is more advantageous to cope with the large state and action spaces \cite{human}. Applying DRL as a solution for radio resource management and core network slicing is investigated in \cite{120}, where a particular scenario with three service types (VoIP, video, URLLC) and hundred users is considered. DRL-based centralized agent for C-RAN slicing is investigated in \cite{115,116}. In \cite{115}, Deep Q-network (DQN) is utilized by the cloud server to optimally manage centralized caching and radio resources and support two transmission-mode network slices, hotspot slice which supports high-transmission-rate users for MBB applications, and vehicle-to-infrastructure slice for delay-guaranteed transmission. C-RAN with single base station is considered by \cite{116}, where Generative-adversarial-network distributed DQN (GAN-DDQN) is examined for dynamic bandwidth slicing among network slices each of which supports users of a particular service type. The dependency of radio resource allocation and the number of slices supported by a single BS is studied by \cite{117}, in which distributed DRL is utilized to achieve optimal and flexible radio resource allocation regardless of the number of slices. \cite{118} utilizes DQN to dynamically select the best slicing configuration in WiFi networks. DRL slicing for dynamic resource reservation is studied in \cite{123}, in which the infrastructure provider assigns the unutilized resources to network slices maintaining a minimum resource requirement and demand in each slice, where DRL then is employed to efficiently manage reserved resources and maximize QoS.

Considering the closest works in the network slicing literature, this paper addresses a more realistic network slicing problem for efficiently allocating edge resources in a diverse IoV and smart city environment. Firstly, dealing with a single fog node as in \cite{access,globalsip,incumbent,117} does not depict the desired network slicing in 5G. A more realistic model should consider a network of multiple coordinated fog nodes, and a comprehensive set of system state variables.

Secondly, the centralized cloud network slicing approach in \cite{113,114,115,116} to manage resource allocation among various network slices is not suitable for delay-sensitive implementations, especially for URLLC-based IoV, V2X, and smart city applications, such as autonomous driving. Whereas, an independent edge DRL agent can avoid large transmission delays and satisfy the desired level of QoS at FNs by closely interacting with the IoV and smart city environment and making local real-time decisions.

Thirdly, the nature of many smart city and IoV applications demands continuous edge capabilities everywhere in the service area (e.g., autonomous vehicles can be anywhere), hence radio, caching and computing resources should be available at the edge. In practice, the demand for delay-sensitive and high-data-rate services will dynamically vary, and as a result the fixed URLLC slice and MBB slice approach in \cite{105,106,111} will cause inefficient utilization of edge resources. For instance, the URLLC slice will be underutilized during light demand for delay-sensitive services. A more flexible network slicing method would efficiently utilize the edge resources while also satisfying the desired QoS. 

Lastly, the hierarchical network slicing architecture and physical resource changes proposed in \cite{111,123} cannot address dynamic environments in a cost-efficient manner. It will be costly for cellular operators and service providers to keep adding or transferring further infrastructural assets, i.e., capital expenditure which includes transceivers (TRX) and other radio resources, computing and signal processing resources such as, BBUs, CPUs and GPUs, as well as caching resources and storage data centers. Such major network changes could be considered part of network expansion plans over time. In this work, we consider a cost-efficient virtual and adaptive network slicing method in F-RAN.

The remainder of the paper is organized as follows. Section \ref{model} introduces the network slicing model. The proposed MDP formulation for the network slicing problem is provided in Section \ref{problem}. Optimal policies and the proposed DRL algorithm are discussed in Section \ref{policy}. Simulation results are presented in Section \ref{simulation}, and the paper is concluded in Section \ref{conclusion}.

\section{Network Slicing Model}
\label{model}
\begin{figure}[!]
\centering
\includegraphics[width=.36\textwidth]{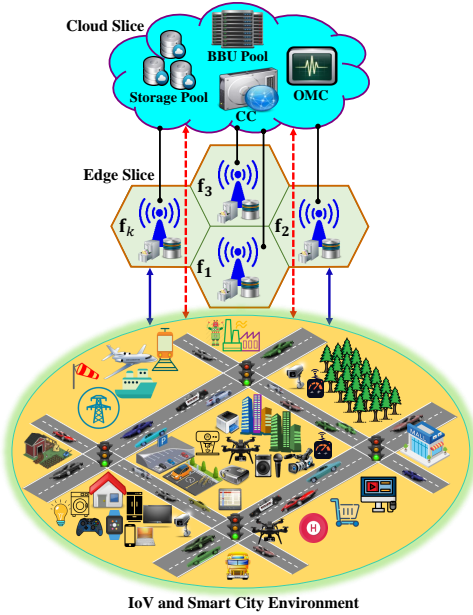}
\caption{Network slicing model. Edge slice is connected to cloud slice through high-capacity fronthaul links represented by solid lines. Solid arrows represent edge service to satisfy QoS, and dashed arrows represent task referral to the cloud-slice to save the limited resources of edge slice.}
\label{f:model}
\end{figure}

We consider the F-RAN network slicing model for IoV and smart city shown in Fig. \ref{f:model}. The two logical network slices, cloud slice and edge slice, support multiple radio access technologies and serve heterogeneous latency needs and resource requirements in dynamic IoV and smart city environments. The hexagonal structure represents the coverage area of fog nodes (FNs) in the edge slice, where each hexagon exemplifies an FN's footprint, i.e., its serving zone. An FN in an edge cluster is connected through extremely fast and reliable optical links with its adjacent FNs whose hexagons have a common side with it. FNs in the edge slice are also connected via high-capacity fronthaul links to the cloud slice which includes a powerful cloud controller (CC) of massive computing capabilities, a pool of huge storage capacity, centralized baseband units (BBUs), and an operations and maintenance center (OMC) which monitors the key performance indicators (KPIs) and generates network reports. To ensure the best QoS for the massive smart city and IoV service requests, especially the URLLC applications and to mitigate the onus on the fronthaul and the cloud, FNs are equipped with computing and processing capabilities to independently deliver network functionalities at the edge of network. However, the edge resources at FNs are limited, and therefore need to be used efficiently. 

In an environment densely populated with low-latency service requests, it is rational for the FNs to route delay-tolerant applications to the cloud and save the limited edge resources for delay-sensitive applications. However, in practice, smart city and IoV environments are dynamic, i.e., a typical environment will not be always densely populated with delay-sensitive applications. A rule-based network slicing policy cannot ensure efficient use of edge resources in dynamic environments as it will under-utilize the edge resources when delay-sensitive applications are rare. On the other hand, a statistical learning policy can adapt its decisions to the changing environment characteristics. Moreover, it can learn to prioritize low-load delay-sensitive applications over high-load delay-sensitive ones. 

We propose using edge controllers (ECs) to efficiently manage the edge resources by enabling cooperation among FNs. In this approach, FNs are grouped in clusters, each of which covers a particular geographical area, and manages the edge resources in that area through a cluster head called EC. The cluster size $k$ is a network design parameter which represents the number of coordinated FNs in an edge cluster. An FN in each cluster is appointed as EC to manage and coordinate edge resources at FNs in the cluster. The EC is nominated by the network designer mainly based on its central geo location among the FNs in the cluster, like $\rf_1$ and $\rf_3$ in Fig. \ref{f:model}. Note that unlike the cloud controller, the edge controller is close to the end users as it is basically one of the FNs in a cluster. Also, the cluster size $k$ is constrained by the neighboring FNs that cover a limited service area such as a downtown, industrial area, university campus, etc.

All FNs in an edge cluster are connected together and with the EC through super-speedy reliable optical links. The EC monitors all individual FN internal states, including resource availability and received service requests, and decides for each service request received by an FN in the cluster. For each received request, the EC chooses one of the three options: serve at the receiving FN (primary FN), serve at a neighboring FN, or serve at the cloud. Each FN in the cluster has a predefined list $\cN_i$ of neighboring FNs, which can help serving a received service request. For instance, $\cC=\{\rf_1,\rf_2,\dots,\overset{\star}{\rf_i},\dots,\rf_k\}$ is an edge cluster of size $k$, where $\overset{\star}{\rf_i}$ denotes the EC which can be any FN in the cluster $\cC$. The network designer needs to define a neighboring list $\cN_i \subseteq \{\cC - \rf_i\}$ for each FN in the cluster. An FN can hand-over service tasks only to its neighbors. Dealing with IoV and smart city service requests, we call the FN which receives a request the primary FN $\hat{\rf}$, and call the FN which actually serves the request utilizing its resources the serving FN $\bar{\rf}$. Depending on the EC decision, the primary FN or one of its neighbors can be the serving FN, or there can be no serving FN (for the decision to serve at the cloud).

An IoV or smart city application attempts to access the network by sending a service request to the primary FN , which is usually the closest FN to the user. The primary FN checks the utility $u \in U=\{1,2,\ldots,u_{max}\}$, i.e., the priority level of executing the service task at the edge, analyzes the task load by figuring the required amount of resources $c \in C=\{1,2,\ldots,c_{max}\}$ and holding time of resources $h \in H=\{1,2,\ldots,h_{max}\}$, and sends the EC the task input $(u_t,c_t,h_t)$ at time $t$.
We consider the resource capacity of the $i^{\text{th}}$ FN $\rf_i \in \cC$ is limited to $\rN_i$ resource blocks. Hence, the maximum number of resource blocks to be allocated for a task is constrained by the FN resource capacity, i.e., $c \le c_{max} \le \rN$. We partition the time into very small time steps $t=1, 2, ...$, and assume a high-rate sequential arrival of IoV and smart city service requests, one task at a time step. ECs should be intelligent to learn how to decide ($\text{which FN to } serve$ or $reject$) for each service request, i.e., how to sequentially allocate limited edge resources, to achieve the objective of efficiently utilizing the edge resources while maximizing the grade-of-service (GoS) defined as the proportion of served high-utility requests to the total number of high-utility requests received. 

A straightforward approach to deal with this network slicing problem is to filter the received service requests by comparing their utility values with a predefined threshold. For instance, consider ten different utilities $u\in \{1,2,3,...,10\}$ for all received tasks in terms of the latency requirement, where $u=10$ represents the highest-priority and lowest-latency task such as the emergency requests from the driver distraction alerting system, and $u=1$ is for the lowest-priority task with highest level of latency such as a service task from smart waste management system. Then, a straightforward non-adaptive solution for network slicing is to dedicate the edge resources to high-utility tasks, such as $u\ge u_h$, and refer to the cloud the tasks with $u<u_h$, where the threshold $u_h$ is a predefined network design parameter. However, such a policy is strictly sub-optimum since the EC will execute any task which satisfies the threshold regardless of how demanding the task load is. For instance, while FNs are busy with serving a few high-utility requests of high load, i.e., low-latency tasks which require large amount of resources $c$ and long holding times $h$, many high-utility requests with low load demand may be missed. In addition, this straightforward policy increases the burden on the cloud unnecessarily, especially when the environment is dominated by low-utility tasks with $u<u_h$. A better network slicing policy would consider the current resource utilization and expected reward of each possible action while deciding, and also adapt to changing utility and load distributions in the environment. To this end, we next propose a Markov Decision Process (MDP) formulation for the considered network slicing problem.

\section{MDP Formulation at EC}
\label{problem}
MDP formulation enables the EC to consider expected rewards of all possible actions in its network slicing decision. Since closed form expressions typically do not exist for the expected reward of each possible action at each system state in a real-world problem, reinforcement learning (RL) is commonly used to empirically learn the optimum policy for the MDP formulation. The RL agent (the EC in our problem) learns to maximize the expected reward by trial and error. That means the RL agent sometimes exploits the best known actions, and sometimes, especially in the beginning of learning, explores other actions to statistically strengthen its knowledge of best actions at different system states. Once, the RL agents learns a optimum policy (i.e., the RL algorithm converges) through managing this exploitation-exploration trade-off, the learned policy can be exploited as long as the environment (i.e., the probability distribution of system state) remains the same. In dynamic IoV and smart city environments, an RL agent can adapt its decision policy to the changing distributions.   

As illustrated in Fig. \ref{f:decision}, for each service request  in an edge cluster at time $t$ from an IoV or smart city application with utility $u_t$, the primary FN computes the number of resource blocks $c_t$ and the holding time $h_t$ which are required to serve the task locally at the edge. Then, the primary FN shares $(u_t, c_t, h_t)$ with the EC , which keeps track of the available resources at all FNs in the cluster. If neither the primary FN nor its neighbors has $c_t$ available resource blocks for a duration of $h_t$, the EC inevitably rejects serving the task at the edge and refers it to the cloud. Note that in the proposed cooperative structure enabled by the EC, such an automatic rejection will be much less frequent compared to the non-cooperative structure considered in \cite{access,globalsip,incumbent,117}, where each FN decides for its resources on its own. If the requested resource blocks $c_t$ for the requested duration $h_t$ are available at the primary FN or at least one of the neighbors, then the EC uses the RL algorithm given in the next section to decide either to serve or reject. In any case, as a result of the taken action $a_t$, the EC will observe a reward $r_t$ and the system state $s_t$ will transition to $s_{t+1}$. We next explain the KPIs in an F-RAN to guide the design of the proposed MDP formulation.

\subsection{Key Performance Indicators}
\label{kpis}

Considering the main motivation behind F-RAN we define the Grade of Service (GoS) as a key performance indicator (KPI). GoS is the proportion of the number of served high-utility service tasks to the total number of high-utility requests in the cluster, and given by
\be
\label{e:gos}
\text{GoS} = \frac{m_h}{M_h} = \frac{\sum_{t=0}^{T-1} \ind{u_t\ge u_h} \ind{a_t \in \{1,2,\dots,k\}} }{\sum_{t=0}^{T-1} \ind{u_t\ge u_h}},
\ee
where $u_h$ is a utility threshold which differentiates the low-latency (i.e., high-utility) tasks such as URLLC from other tasks, and $\ind{\cdot}$ is the indicator function taking the value $1$ when its argument is true, and $0$ otherwise. 

Naturally, the network operator would want the edge resources to be efficiently utilized. Hence, the average utilization of edge resources over a time period $T$ gives another KPI:
\be
\label{e:utilization}
\text{Utilization} = \frac{1}{T} \sum_{t=0}^{T-1} \frac{\sum_{i=1}^{k}{b_i}_t}{\sum_{i=1}^{k}\rN_i},
\ee
where ${b_i}_t$ and $\rN_i$ are the number of occupied resources at time $t$, and the resource capacity of the $i^{\text{th}}$ FN in the cluster, respectively. Another KPI to examine the EC performance is cloud avoidance which is given by the proportion of all IoV and smart city service requests that are served by FNs in the edge cluster to all requests received. Cloud avoidance is reported over a period of time $T$, and it is given by
\be
\label{e:cloud_avoidance}
\text{Cloud Avoidance} = \frac{m}{M} = \frac{\sum_{t=0}^{T-1} \ind{a_t \in \{1,2,\dots,k\}}}{M},
\ee
where $m=m_h+m_l$ is the number of high-utility and low-utility served requests at the edge cluster, and $M=M_h+M_l$ is the total number of high-utility and low-utility received requests. Note that $M-m$ is the portion of IoV and smart city service tasks which is served by the cloud, and one of the objectives of F-RAN is to lessen this burden especially during busy hours. Cloud avoidance shows a general overview about the contribution of edge slice to share the load. It gives a similar metric as resource utilization, which is more focused on resource occupancy rather than dealing with service requests in general. While we use the resource utilization together with GoS to define an overall performance metric below, cloud avoidance is still used as a performance evaluation metric in Sec. \ref{simulation}.

To evaluate the performance of an EC over a particular period of time $T$, we consider the weighted sum of the main two KPIs, the GoS and edge-slice average resource utilization as
\be
\label{e:performance}
\text{Performance} = \omega_g \text{GoS} + \omega_u \text{Utilization}.
\ee

\subsection{MDP Formulation}
\label{mdp}
An MDP is defined by the tuple $(\cS,\cA,P_a,R_a,\gamma)$, where $\cS$ is the set of states, i.e., $s_t \in \cS$, $\cA$ is the set of actions, i.e., $a_t \in \cA=\{1,2,\dots,k,k+1\}$, $P_a(s,s')=P(s_{t+1}=s'|s_t=s,a_t=a)$ is the transition probability from state $s$ to $s'$ when action $a$ is taken, $R_a(s,s')$ is the reward received by taking action $a$ in state $s$ which ends up in state $s'$, i.e., $r_t \in R_a(s,s')$, and $\gamma \in [0,1]$ is the discount factor in computing the return which is the cumulative reward
\be
\vspace{-1mm}
\label{e:G}
G_t = r_t+\gamma r_{t+1}+\gamma^2 r_{t+2}+...=\sum_{j=0}^{\infty} \gamma^j r_{t+j}.
\ee
$\gamma$ represents how much weight is given to the future rewards compared to the immediate reward. For $\gamma=1$, future rewards are of equal importance as the immediate reward, whereas $\gamma=0$ completely ignores future rewards. The objective in MDP is to maximize the expected cumulative reward starting from $t=0$, i.e., $\max\limits_{\{a_t\}} \Exp[G_0 | s_0]$, where $G_t$ is given by \eqref{e:G}, by choosing the actions $\{a_t\}$. 

\begin{figure}[!]
\centering
\includegraphics[width=.47\textwidth]{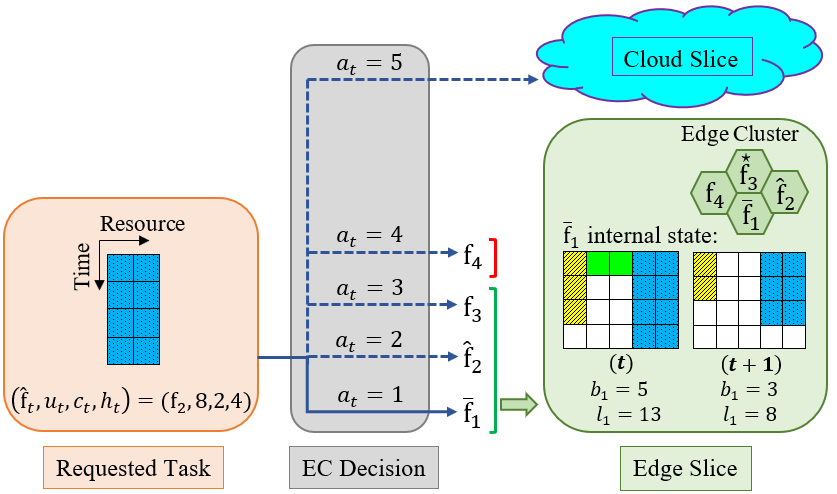}
\caption{EC decision for a sample service request received by $\rf_2$, and the internal state of the serving FN $\rf_1$ with $\rN_1=5$ and $h_{max}=4$ (see \eqref{e:s}). The edge cluster size is $k=4$ and $\rf_3$ is the EC.}
\label{f:decision}
\end{figure}

Before explaining the (state, action, reward) structure in the proposed MDP, let us first define the task load $L_t$ as the number of resource blocks required to execute a task completely,
\be
\label{e:load}
L = c \times h,
\ee
and similarly the existing load ${l_i}_t$ of FN $i$ as the number of already allocated resource blocks (see Fig. \ref{f:decision}).  

\begin{itemize}
\item \textbf{State:} We define the system state in a cluster of size $k$ at any time $t$ as
\be
\label{e:s}
s_t = ({b_1}_t,{l_1}_t,{b_2}_t,{l_2}_t,\dots,{b_k}_t,{l_k}_t,\hat{\rf}_t,u_t,c_t,h_t),
\ee
where ${b_i}_t$ denotes the number of resource blocks in use at FN $i$ at time $t$. Note that ${b_i}_{(t+1)}, {l_i}_{(t+1)}$ and in turn the next state $s_{t+1}$ are independent of the past values given the current state $s_t$, satisfying the Markov property $P(s_{t+1}|s_0, s_1, s_2, ..., s_t,a_t) = P(s_{t+1}|s_t,a_t)$.
\item \textbf{Action:} The EC decides, as shown in Fig. \ref{f:decision}, for each service request by taking an action $a_t \in \cA=\{1,2,\dots,k,k+1\}$, where $a_t=i\in \{1,2,\dots,k\}$ means $serve$ the requested task at the $i^{\text{th}}$ FN in the cluster, $\rf_i \in \cC=\{\rf_1,\rf_2,\dots,\rf_k\}$, whereas $a_t = k+1$ means to $reject$ the job and refer it to the cloud. Note that for a request received by $\rf_i$, the feasible action set is a subset of $\cA$ consisting of $\rf_i$, its neighbors $\cN_i$, and the cloud. Fig. \ref{f:decision} illustrates the decision of the EC for a sample service request received by $\rf_2$ at time $t$ in an edge cluster with $k\!=\!4$ FNs. Note that in this example the action $a_t\!=\!4$ is not feasible as $\rf_4\! \notin \! \cN_2$, and the EC took the action $a_t\!=\!1$, which means serve the task by $\rf_1$. Hence, $\rf_1$ started executing the task at $t$ while another two tasks (striped yellow and green) are in progress. At $t\!+\!1$, two resource blocks are released as the job in clear-green is completed. Note that resource utilization of $\rf_1$ decreased from $100\%$ at $t$, i.e., internal busy state with ${b_1}_t\!=\!5$, to $60\%$ at $t\!+\!1$.
\item \textbf{Reward:} 
In general, a proper rewarding system is crucial for an RL agent to learn the optimum policy of actions that maximizes the KPIs. The EC RL agent collects an immediate reward $r_t \in R_a(s,s')$ for taking action $a$ at time $t$ from state $s$ which ends in state $s'$ in the next time step $t+1$. We define the immediate reward
\be
\label{e:r}
r_t = r_{(a_t,u_t)} \pm r_{L_t}
\ee
using two components. The first term $r_{(a_t,u_t)} \in \{r_{sh}, r_{sl}, r_{rh}, r_{rl}, r_{bh}, r_{bl}\}$ corresponds to the reward portion for taking an action $a\in \{1,2,\dots,k,k+1\}$ when a request of specific $u$ is received, and the second term
\be
\label{e:r_l}
r_{L_t} = c_{max} \times h_{max}+1-L_t,
\ee
considers the reward portion for handling the new job load $L_t=c_t \times h_t$ of a requested task. For instance, serving low-load task such as $L=3$ is awarded more than serving a task with $L=18$. Similarly, rejecting a low-load task such as $L=3$ should be more penalized, i.e., negatively rewarded especially when $u \ge u_h$, than rejecting a task with the same utility and higher load such as $L=18$. The two reward parts are added when $a_t=serve$, and subtracted if $a_t=reject$. 
We define six different reward-component $r_{(a,u)} \in \{r_{sh}, r_{sl}, r_{rh}, r_{rl}, r_{bh}, r_{bl}\}$, where $r_{sh}$ is the reward for serving a high-utility request, $r_{sl}$ is the reward for serving a low-utility request, $r_{rh}$ is the reward for rejecting a high-utility request, $r_{rl}$ is the reward for rejecting a low-utility request, $r_{bh}$ is the reward for rejecting a high-utility request due to being busy, and $r_{bl}$ is the reward for rejecting a low-utility request due to being busy. Note that having a separate reward for rejecting due to a busy state makes it easier for the RL agent to differentiate between similar states for the $reject$ action. A request is determined as high-utility or low-utility based on the threshold $u_h$, which is a design parameter that depends on the level of latency requirement in an IoV and smart city environment. 
\end{itemize}


\section{Optimal Policies and DQN}
\label{policy}
The state value function $V(s)$ represents the long-term value of being in a state $s$. That is, starting from state $s$ how much value on average the EC will collect in the future, i.e., the expected total discounted rewards from that state onward. Similarly, the action-value function $Q(s,a)$ tells how valuable it is to take a particular action $a$ from the state $s$. It represents the expected total reward which the EC may get after taking the particular action $a$ from the state $s$ onward. The state-value and the action-value functions are given by the Bellman expectation equations \cite{sutton},
\begin{align}
\label{e:V}
V(s) &= \Exp[G_t|s]= \Exp[r_t+\gamma V(s') | s], \\
\label{e:Q}
Q(s,a) &= \Exp[G_t|s,a]= \Exp[r_t+\gamma Q(s',a') | s,a],
\end{align}
where the state value $V(s)$ and the action value $Q(s,a)$ are recursively presented in terms of the immediate reward $r_t$ and the discounted value of the successor state $V(s')$ and the successor state-action $Q(s',a')$, respectively.

Starting at the initial state $s_0$, the EC objective can be achieved by maximizing the expected total return $V(s_0)=\Exp[G_0 | s_0]$ over a particular time period $T$. To achieve this goal, the EC should learn an optimal decision policy to take proper actions. However, considering the large dimension of sate space (see \eqref{e:s}) and the intractable number of state-action combinations, it is infeasible for RL tabular methods to keep track of all state-action pairs and continuously update the corresponding $V(s)$ and $Q(s,a)$ for all combinations in order to learn the optimal policy. Approximate DRL methods such as DQN is a more efficient alternative for the high-dimensional EC MDP to quickly learn an optimal decision policy to take proper actions, which we discuss next.

A policy $\pi$ is a way of selecting actions. It can be viewed as a mapping from states to actions as it describes the set of probabilities for all possible actions to select from a given state, i.e., $\pi=\{P(a|s)\}$. A policy helps in estimating the value functions in \eqref{e:V} and \eqref{e:Q}. $\pi_1$ is said to be better than another policy $\pi_2$ if the state value function following $\pi_1$ is greater than that following $\pi_2$ for all states, i.e., ${\pi_1>\pi_2}$ if ${V_{\pi_1}(s)>V_{\pi_2}(s)}, \forall {s \in \cS}$. A policy $\pi$ is said to be optimal if it maximizes the value of all states, i.e., ${\pi^*=\arg \max\limits_\pi   V_{\pi}(s)}, {\forall s \in \cS}$.
Hence, to solve the considered MDP problem, the DRL agent needs to find the optimal policy through finding the optimal state-value function ${V^*(s)=\max\limits_\pi V_{\pi}(s)}$, which is similar to finding the optimal action-value function ${Q^*(s,a)=\max\limits_\pi Q_{\pi}(s,a)}$ for all state-action pairs.
From \eqref{e:V} and \eqref{e:Q}, we can write the Bellman optimality equations for $V^*(s)$ and $Q^*(s,a)$ as,
\begin{align}
\vspace{-6mm}
\label{e:V^*}
V^*(s) &= \max\limits_{a\in \cA} Q^*(s,a)=\max\limits_{a\in \cA} \Exp[r_t+\gamma V^*(s') | s,a], \\
\label{e:Q^*}
Q^*(s,a) &= \Exp[r_t+\gamma \max\limits_{a'\in \cA} Q^*(s',a')|s,a].
\end{align}
The expression of optimal state-value function $V^*(s)$ greatly simplifies the search for optimal policy as it subdivides the targeted optimal policy into local actions: take an optimal action $a^*$ from state $s$ which maximizes the expected immediate reward followed by the optimal policy from successor state $s'$. Hence, the optimal policy is simply taking the best local actions from each state considering the expected rewards. Dealing with $Q^*(s,a)$ to choose optimal actions is even easier, because with $Q^*(s,a)$ there is no need for the EC to do the one-step-ahead search and instead it picks the best action that maximizes $Q^*(s,a)$ at each state. The optimal action for each state $s$ is given by
\begin{equation}
\label{e:a^*}
a^*=\arg \max\limits_{a\in \cA} Q^*(s,a)=\arg \max\limits_{a\in \cA} \Exp[r_t+\gamma V^*(s')|s,a].
\end{equation}

The optimal policy can be learned by solving the Bellman optimality equations \eqref{e:V^*} and \eqref{e:Q^*} for $a^*$. This can be done for tractable number of states by estimating the optimal value functions using tabular solution methods such as dynamic programming, and model-free RL methods which include Monte Carlo, SARSA, expected SARSA, and Q-Learning (QL) \cite{access}. However, for high-dimensional state space, such as ours given in \eqref{e:s}, tabular methods are not tractable in terms of computational and storage complexity. Deep RL (DRL) methods address the high-dimensionality problem by approximating the value functions using deep neural networks (DNN).

\begin{figure}[!]
\centering
\includegraphics[width=.45\textwidth]{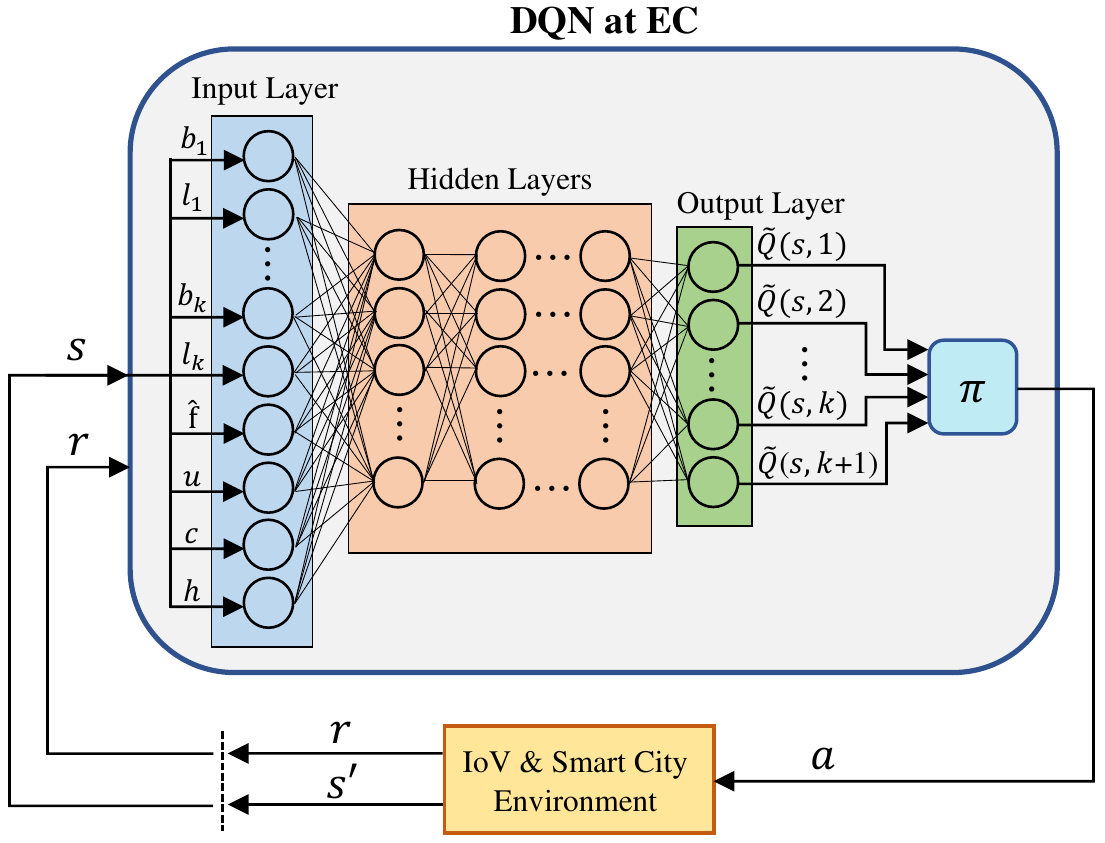}
\caption{The interaction of the DQN-based EC with the IoV and smart city environment. Given the EC input state $s=(b_1,l_1,\dots,b_k,l_k,\hat{\rf},u,c,h)$, the DQN agent predicts the action-value functions and follows a policy $\pi$ to take an action $a$ which ends up in state $s'$, and collects a reward $r$ accordingly.}
\label{f:dqn}
\end{figure}

Deep Q-Network (DQN) is a powerful DRL method for addressing RL problems with high-dimensional input states and output actions \cite{human}. DQN extends QL to high-dimensional problems by using DNN to approximate the action-value functions without keeping a $Q$-table to store and update the $Q$-values for all possible state-action pairs as in QL. Fig. \ref{f:dqn} demonstrates the DQN method for EC in the network slicing problem, in which the DQN agent at EC learns about the IoV and smart city environment by interaction. The DQN agent is basically a DNN that consists of an input layer, hidden layers, and an output layer. The number of neurons in the input and output layers is equal to the state and action dimensions, respectively, whereas the number of hidden layers and the number of neurons in each hidden layer are design parameters to be chosen. Feeding the current EC state $s$ to the DNN as an input and regularly updating its parameters, i.e., the weights of all connections between neurons, DNN is able to predict the $Q$-values at the output for a given input state. The DRL agent at EC sometimes takes random actions to explore new rewards, and at other times exploits its experience to maximize the discounted cumulative rewards over time and keeps updating the DNN weights. Once the DNN weights converge to the optimal values, the agent learns the optimal policy for taking actions in the observed environment.

For a received service request, if the requested resources are affordable, i.e., ${c_t \le (\rN_i-{b_i}_t)}$ for any ${\rf_i \in \{\hat{\rf}_i,\cN_i\}}$, the EC makes a decision whether to $serve$ the request by the primary FN or one of its neighbors, or $reject$ and refer it to the cloud. From \eqref{e:a^*}, the optimal action at state $s$ is given by,

\begin{equation}
\label{e:aa^*}
a^*=
\begin{cases}
\begin{aligned}
i \in \tilde{\cA}& \hspace{6pt} \text{if}\  Q^*(s,i)=\max\limits_{a \in \{\tilde{\cA}, k+1\}} Q(s,a),\\
k+1& \hspace{6pt} \text{otherwise},
\end{aligned}
\end{cases} 
\end{equation}
where $\tilde{\cA}$ denotes the set of possible $serve$ actions to execute the service task by $f_i \in \{\hat{f}_i, \cN_i\}$. The procedure to learn the optimal policy from the IoV and smart city environment using the model-free DQN algorithm is given in Algorithm \ref{a:dqn}.

\begin{algorithm}
\caption{Learning Optimum Policy Using DQN }
\label{a:dqn}
\begin{algorithmic} [1]
\STATE Select: \{$\gamma, \epsilon\} \in [0, 1]$, $\rho \in (0, 1]$, $n \in \{1,2,\dots,D\}$;
\STATE Create DNN model and target model with weights $\rw$ and $\hat{\rw}$, respectively;
\STATE Initialize: $\rw$, $\hat{\rw}$, the replay memory $\M$ with size $D$;
\STATE Initialize: $s$;
\FOR{$t=0,1,2,\dots,T$}
\STATE Take action $a_t$ according to $\pi=\epsilon$-greedy, and observe $r_t$ and $s_{t+1}$;
\STATE Append the observation $(s_t,a_t,r_t,s_{t+1})$ to $\M$;
\STATE $s \leftarrow s_{t+1}$;
\STATE Sample a random minibatch of $n$ observations from $\M$;
\FOR{$j=1,2,\dots,n$}
\STATE Predict $\hat{Q}_j(s_j|\hat{\rw})$;
\STATE $\hat{Q}_j[a_j] = 
\begin{cases} 
\begin{aligned} 
&r_j \hspace{6pt} \text{if}\  t+1=T,\\ 
&r_j+\gamma \max\limits_{a'} \hat{Q}_j(s',a'|\hat{\rw}) \hspace{6pt} \text{otherwise}. 
\end{aligned} 
\end{cases}$ 
\STATE Fit the DNN model for $(s_j,\hat{Q}_j)$ by applying gradient descent step on ${\left(\hat{Q}_j - \tilde{Q}_j\right)}^2$ with respect to $\rw$;
\ENDFOR
\IF{$(t \ \text{mod} \ \tau) = 0$}
\STATE $\hat{\rw} \leftarrow \rho \rw +(1-\rho) \hat{\rw}$;
\ENDIF
\IF{$\rw$ converges}
\STATE $\rw^* \leftarrow \rw$;
\STATE \textbf{break}
\ENDIF
\ENDFOR
\STATE Use $\rw^*$ to estimate ${Q^*(s,a)}$ required for $\pi^*$ using \eqref{e:aa^*}.
\end{algorithmic}
\end{algorithm}

Algorithm \ref{a:dqn} shows how the EC learns the optimal policy $\pi^*$ for the considered MDP. It requires the EC design parameters $k$, $\cN$, $\rN$, and $u_h$, and selecting the DNN hyper parameters $\gamma$, the target update rate $\rho$, the probability $\epsilon$ of making a random action for exploration, the replay memory capacity $D$ to store the observations $(s,a,r,s')$, the minibatch size $n$ of samples used to train the DNN model and update its weights $\rw$, and the data of the IoV and smart city users $u$, $c$, $h$. Note that $u$, $c$ and $h$ can be real data from the IoV and smart city environment, as well as from simulations if the probability distributions are known. The DNN target model at line 2 is used to stabilize the DNN model by reducing the correlation between the action-values $Q(s,a)$ and the targets ${r+\gamma \max\limits_{a'} Q(s',a')}$ through only periodical updates of the target model weights $\hat{\rw}$. In each iteration, we take an action and observe the collected reward and the successor state. Actions are taken according to a policy $\pi$ such as the $\epsilon$-greedy policy in which a random action with probability $\epsilon$ is taken to explore new rewards, and an optimal action (see \eqref{e:aa^*}) is taken with probability $(1-\epsilon)$ to maximize the rewards. Model is trained using experience replay as shown in lines 9-14. At line 9, a minibatch of $n$ random observations is sampled from $\M$. The randomness in selecting samples eliminates correlations in the observations to avoid model overfitting. At line 11, we estimate the output vector $\hat{Q}$ of the target model for a given input state $s$ in each experience sample using the target model weights $\hat{\rw}$. $\tilde{Q}$ and $\hat{Q}$ are the predicted vectors of the ${k\!+\!1}$ $Q$-values for a given state $s$ with respect to $\rw$ and $\hat{\rw}$, respectively. The way to update the action-value for sample $j$ is shown at line 12, where the element value $\hat{Q}_j[a_j]$ is replaced with the immediate reward $r_j$ if state is terminal, i.e., ${t\!=\!T\!-\!1}$, or with the collected immediate reward and a discounted value of the maximum action-value considering all possible actions which can be taken from the state at ${t\!+\!1}$ if ${t\!<\!T\!-\!1}$. At line 13, we update the model weights $\rw$ by fitting the model for the input states and the updated predicted outputs. A gradient decent step is applied to minimize the squared loss ${\left(\! \hat{Q}_j\!-\! \tilde{Q}_j\!\right)}^2$ between the target and the model predictions. The target model weights are periodically updated every $\tau$ time steps as shown at line 16, where the update rate $\rho$ exemplifies how much we believe in our experience. The algorithm stops when the DNN model weights $\rw$ converge. The converged values are then used to determine optimal actions, i.e., $\pi^*$ as in \eqref{e:aa^*}.
\section{Simulations}
\label{simulation}
We next provide simulation results to evaluate the performance of the proposed network slicing approach in dynamic IoV and smart city environments. We compare the DRL algorithm given in Algorithm \ref{a:dqn} with the serve-all-utilities (SAU) algorithm in which the EC serves all coming tasks when requested resources are available, serve-high-utilities (SHU) algorithm where the EC filters high-utility requests and serve them if the available resources are enough, and the QL algorithm independently running at each FN following a local version of our MDP formulation \cite{access}. The QL algorithm at each FN corresponds to the non-cooperative scenario, hence this comparison will help evaluate the importance of cooperation among FNs. In the non-cooperative scenario, each FN operates as a standalone entity with no neighbors to handover tasks when busy, and no EC to manage the edge resources.
\begin{table}[!]
\caption{Utility distributions corresponding to a variety of latency requirements of IoV and smart city applications in various environments}
\label{t:env}
\begin{center}
\begin{tabular}{| c| c| c| c| c| c| }
\hline
$$ & $\mathcal{E}_1$ & $\mathcal{E}_2$ & $\mathcal{E}_{3}$ & $\mathcal{E}_{4}$ & $\mathcal{E}_{5}$\\[.1mm]
\hline
$P(u=1)$ &  $0.015$ &  $0.012$ &  $0.008$ &  $0.004$ &  $0.001$\\
$P(u=2)$ &  $0.073$ &  $0.058$ &  $0.038$ &  $0.019$ &  $0.004$\\
$P(u=3)$ &  $0.365$ &  $0.288$ &  $0.192$ &  $0.096$ &  $0.019$\\
$P(u=4)$ &  $0.292$ &  $0.230$ &  $0.154$ &  $0.077$ &  $0.015$\\
$P(u=5)$ &  $0.205$ &  $0.162$ &  $0.108$ &  $0.054$ &  $0.011$\\
$P(u=6)$ &  $0.014$ &  $0.071$ &  $0.142$ &  $0.214$ &  $0.271$\\
$P(u=7)$ &  $0.013$ &  $0.064$ &  $0.129$ &  $0.193$ &  $0.244$\\
$P(u=8)$ &  $0.011$ &  $0.057$ &  $0.114$ &  $0.171$ &  $0.217$\\
$P(u=9)$ &  $0.009$ &  $0.043$ &  $0.086$ &  $0.129$ &  $0.163$\\
$P(u=10)$ &  $0.003$ &  $0.015$ &  $0.029$ &  $0.043$ &  $0.055$\\
$P(u\ge u_h=8)$ &  $2.3\%$ &  $11.5\%$ &  $22.9\%$ &  $34.3\%$ & $43.5\%$\\
$\bar{u}$ &  $3.82$ &  $4.589$ &  $5.55$ &  $6.5$ & $7.27$\\
\hline
\end{tabular}
\end{center}
\end{table}
\begin{table}[!]
\caption{Simulation Setup}
\label{t:sim}
\begin{center}
\begin{tabular}{c @{\ \ } l @{\ \ } c}
\hline\hline
Parameter &\multicolumn{1}{c}{Description} & Value\\
\hline
$\rN$ &  FN resource capacity&  $7$\\
$C$ &  set of possible resource blocks &  $\{1,2,3,4\}$\\
$H$ &  set of possible holding times &  $5$$\times$$\{1,2,3,4,5,6\}$\\
$\omega_g$ &weight for GoS & $\{0.7, 0.5, 0.3\}$\\
$\omega_u$ &weight for resource utilization & $\{0.3, 0.5, 0.7\}$\\
$u_h$ &  threshold for a ``high-utility" &  $8$\\
$D$ &  capacity of DNN replay memory &  $2000$\\
$\gamma$  &reward discount factor &  $0.9$\\
$\alpha$    &learning rate &  $0.01$\\
$\epsilon$ &probability of random action &  $1.0$ with $0.9995$ decay\\
$n$ &  batch size &  $32$\\
$\tau$ &  $\hat{\rw}$ update interval  &  $1000$\\
$\rho$ &  $\hat{\rw}$ update rate &  $0.2$\\
\hline\hline
\end{tabular}
\end{center}
\end{table}
\begin{figure}[!]
\centering
\includegraphics[width=.15\textwidth]{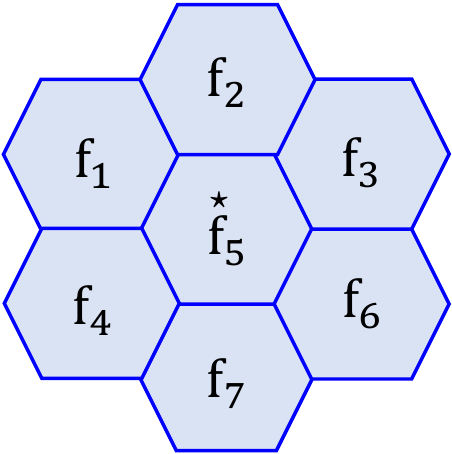}
\caption{The structure of the edge cluster considered in the simulations. The neighboring lists $\cN_1,\cN_2,\dots,\cN_7$ include the adjacent FNs only.}
\label{f:cluster}
\end{figure}
\begin{table}[!]
\caption{Considered Rewarding Systems}
\label{t:r}
\begin{center}
\begin{tabular}{c @{\ \ }c @{\ \ \ }c @{\ \ \ }c @{\ \ }c @{\ \ }c}
\hline\hline
Scenario &$\omega_g$ &$\omega_u$ &$R$ &$\{r_{sh},\ r_{rh},\ r_{bh},\ r_{sl},\ r_{rl},\ r_{bl}\}$ &$r_L$\\
\hline
1 &$0.7$ &$0.3$ &$R_1$ &$\ \{24,-12,-12,\ -3,\ \ \ 3,\ \ \ 12\}$ &(see \eqref{e:r_l})\\
2 &$0.5$ &$0.5$ &$R_2$ &$\ \{24,-12,-12,\ \ \ 0,\ \ \ 0,\ \ \ 12\}$ &(see \eqref{e:r_l})\\
3 &$0.3$ &$0.7$ &$R_3$ &$\ \{50,-50,-50,\ \ 50,-50,-25\}$ &$0$\\
\hline\hline
\end{tabular}
\end{center}
\end{table}
\begin{figure}[!]
\centering
\includegraphics[width=.45\textwidth]{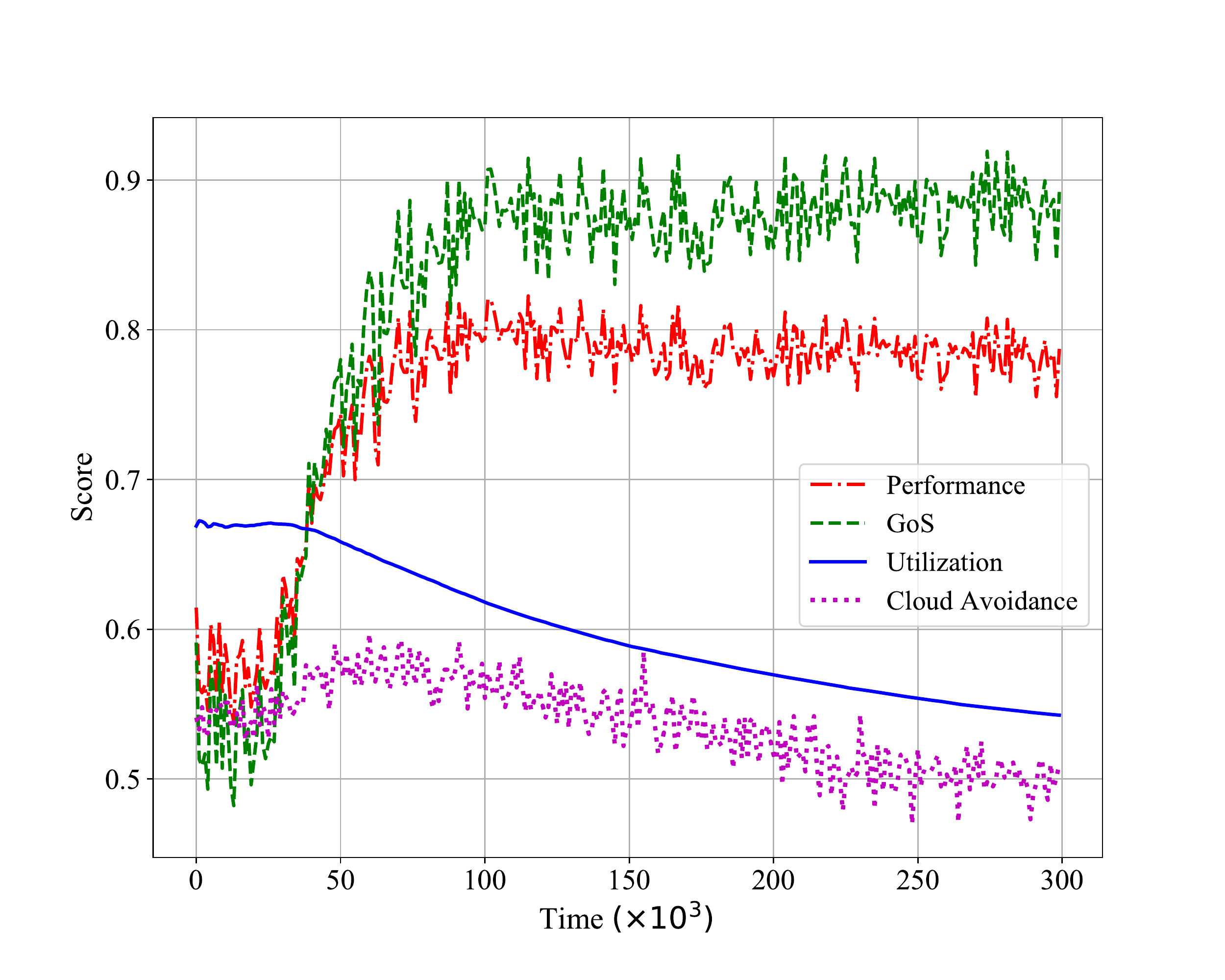}
\caption{The performance and main network KPIs for DQN-based EC while learning the optimum policy in the IoV and smart city environment $\cE_3$ under scenario 1 of Table \ref{t:r}.}
\label{f:convergence}
\end{figure}
\begin{figure}[!]
\centering
\includegraphics[width=.39\textwidth]{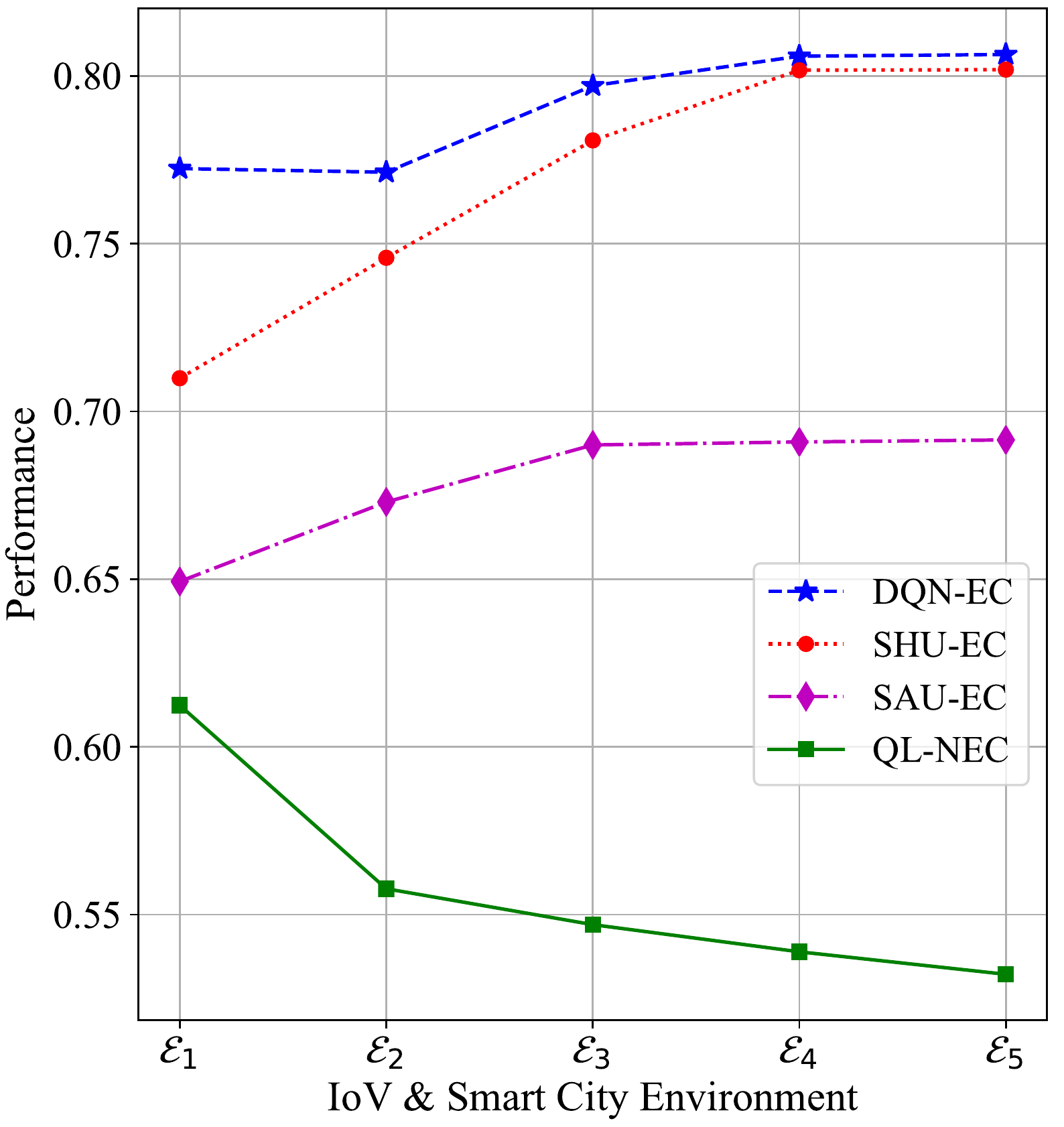}
\caption{The performance of the edge slice when the EC applies the DRL Algorithm \ref{a:dqn}, SAU and SHU for the coordinated FNs, and the uncoordinated QL based FNs case with NEC. Considering scenario 1 in Table \ref{t:r} with $\omega_{g}=1-\omega_{u}=0.7$.}
\label{f:perf70}
\end{figure}
\begin{figure}[t]
\centering
\includegraphics[width=.39\textwidth]{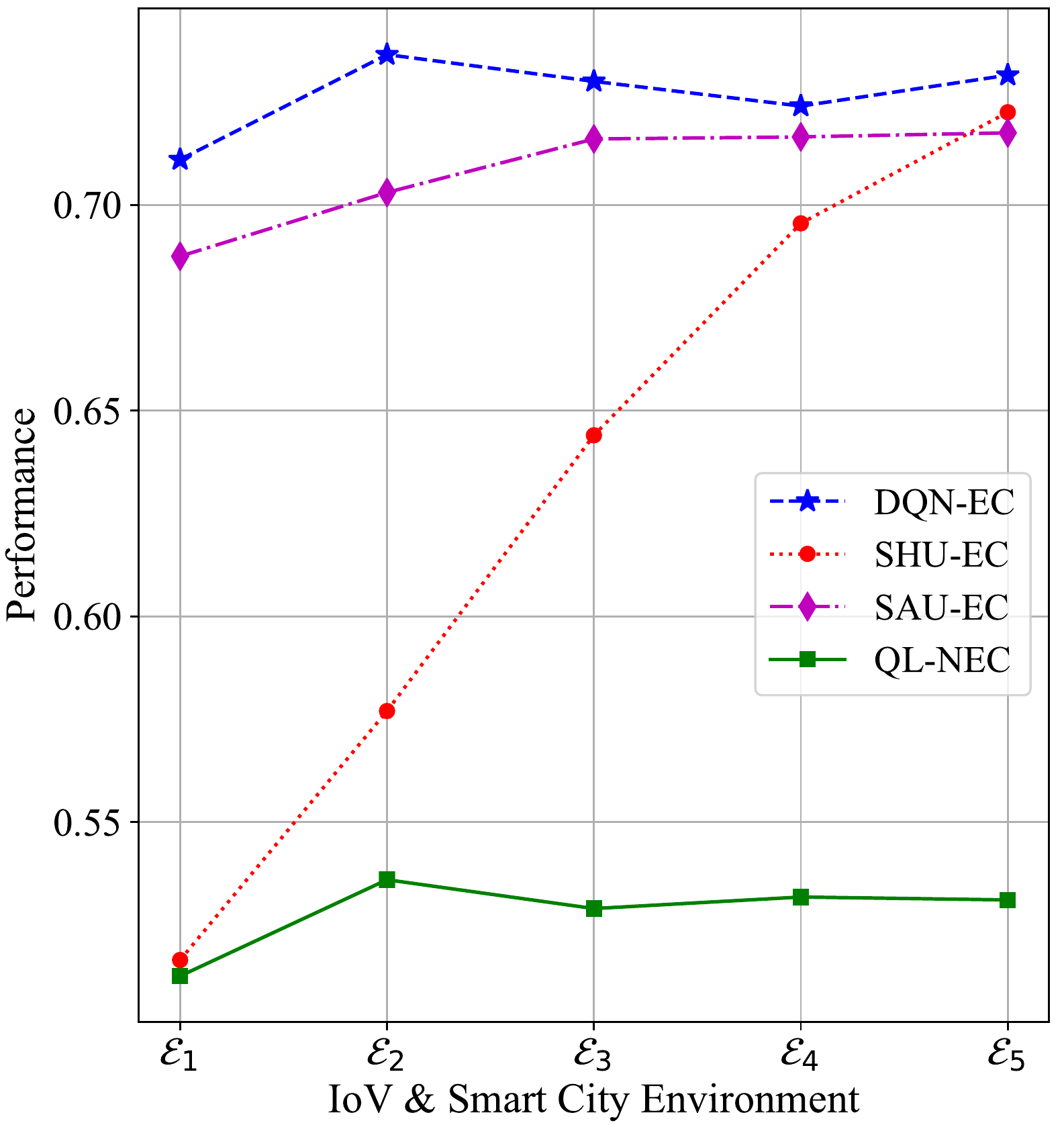}
\caption{The performance of the edge slice when the EC applies the DRL Algorithm \ref{a:dqn}, SAU and SHU for the coordinated FNs, and the uncoordinated QL based FNs case with NEC. Considering scenario 2 in Table \ref{t:r} with $\omega_{g}=1-\omega_{u}=0.5$.}
\label{f:perf50}
\end{figure}
\begin{figure}[t]
\centering
\includegraphics[width=.39\textwidth]{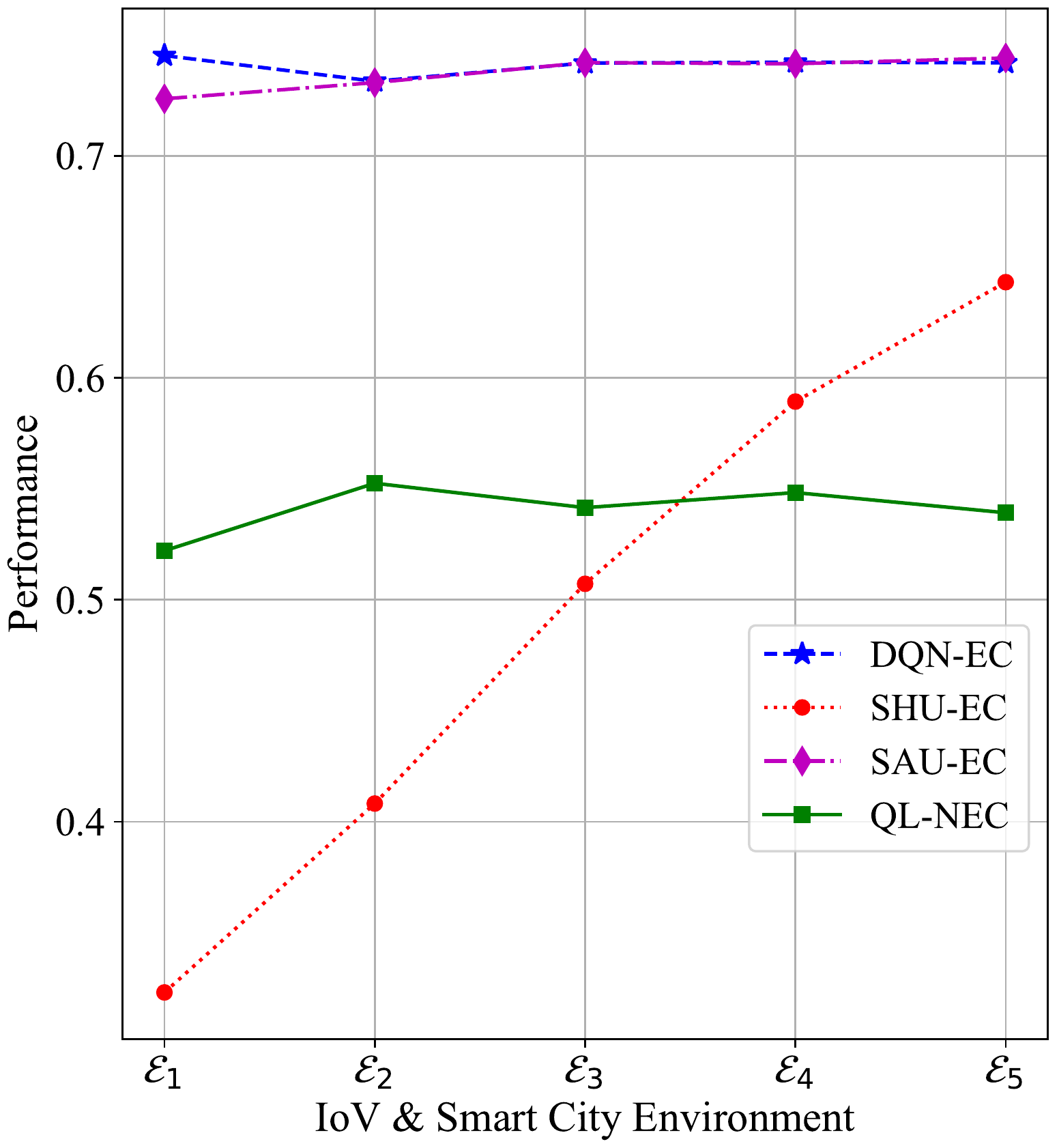}
\caption{The performance of the edge slice when the EC applies the DRL Algorithm \ref{a:dqn}, SAU and SHU for the coordinated FNs, and the uncoordinated QL based FNs case with NEC. Considering scenario 3 in Table \ref{t:r} with $\omega_{g}=1-\omega_{u}=0.3$.}
\label{f:perf30}
\end{figure}

\subsection{Simulation Environments}
\label{env}
We evaluate the performances in various IoV and smart city environments with different compositions of user utilities. Specifically, we consider $10$ utility classes that represent different latency requirements to exemplify the variety of IoV and smart city applications in an F-RAN setting. By changing the distribution of utility classes we generate $5$ IoV and smart city environments as summarized in Table \ref{t:env}. Higher density of high-utility applications makes the IoV and smart city environment richer in terms of URLLC applications. Denoting an IoV and smart city environment of a particular utility distribution with $\cE$, we show in Table \ref{t:env} the statistics of $\cE_1$, $\cE_2$, $\cE_3$, $\cE_4$, and $\cE_5$. The probabilities in the first 10 rows in Table \ref{t:env} present detailed information about the proportion of each utility class in the environment corresponding to the latency requirement of diverse IoV and smart city applications. The last two rows interpret the quality or richness of IoV and smart city environments, where $\bar{u}$ is the mean of utilities in an environment, and $P(u\ge u_h)$ is the percentage of high-utility population. We started with a general environment given by $\mathcal{E}_3$ for the following IoV and smart city applications corresponding to the utility values $1, 2, \ldots, 10$, respectively:  smart lighting and automation of public buildings, air quality management and noise monitoring, smart waste management and energy consumption management, smart parking assistance, in-vehicle audio and video infotainment, driver authentication service, structural health monitoring, safe share rides, smart amber alerting system and AI-driven and video-analytics tracking services, driver distraction alerting system and autonomous driving. Then, we changed the utility distribution to obtain the other environments.

\subsection{Simulation Parameters}
\label{parameters}
The simulation parameters used in this section are summarized in Table \ref{t:sim}. We consider an edge cluster of size $k=7$, where each FN has a computing and processing resource capacity of seven resource blocks, i.e., $N=7$. The central FN $\rf_5$ acts as the EC, and the neighboring relationships are shown in Fig. \ref{f:cluster}. In a particular IoV and smart city environment $\cE$, the threshold that defines ``high utility" is set to $u_h=8$, i.e., $u\in \{8, 9, 10\}$ is a high-utility application with higher priority for edge service. To make the resource allocation of the network slicing problem more challenging, we consider a request arrival rate of at least five times the task execution rate, i.e., holding times increment by five times the arrival interval. The probabilities of $c\in C=\{1,2,3,4\}$ are $0.1$, $0.2$, $0.3$, and $0.4$, respectively, whereas the probabilities of $h\in H=\{5,10,15,20,25,30\}$ are $0.05$, $0.1$, $0.1$, $0.15$, $0.2$, and $0.4$, respectively.

We consider a fully connected DNN structure for DQN with an input layer of 18 neurons, 2 hidden layers of 64 and 24 neurons, respectively, and an 8-neuron output layer. Linear activation function is used at the output layer and ReLU activation is considered for the other layers. The Huber loss function and the RMSprop optimizer are considered with $0.01$ learning rate, $10^{-4}$ learning decay, and momentum of $0.9$. The $\epsilon$-greedy policy is adopted in DNN training where $\epsilon$ starts at $1.0$ for $10\%$ of the time in training and then decays at a rate of $0.9995$ to a minimum value of $10^{-3}$ to guarantee enough exploration over time.

We examine the KPIs explained in Sec. \ref{kpis}, GoS, resource utilization, cloud avoidance, as well as the overall performance (see \eqref{e:performance}-\eqref{e:cloud_avoidance}) considering the three scenarios shown in Table \ref{t:r} with the weights $\omega_g=1-\omega_u=0.7$, $\omega_g=\omega_u=0.5$, and $\omega_g=1-\omega_u=0.3$. Each scenario in Table \ref{t:r} represents a new problem, hence the rewarding systems $R_1$, $R_2$, and $R_3$ are chosen to facilitate learning the optimal policy in each scenario. The two reward components, $r_{(a,u)}\in\{r_{sh},r_{rh},r_{bh},r_{sl},r_{rl},r_{bl}\}$ and $r_L$ for each rewarding system are provided in Table \ref{t:r}. Note that unlike $R_2$ and $R_3$, $R_1$ encourages rejecting low-utility requests with higher loads to accommodate the performance requirement of scenario 1, which puts higher weight on GoS with $\omega_g=0.7$. On the other hand, $R_3$ promotes serving regardless of the request utility and the task load as the performance in scenario 3 is in favor of achieving higher resource utilization with $\omega_u=0.7$.

\begin{figure*}
	\centering
	\begin{subfigure}{.3\textwidth}
		\includegraphics[width=\linewidth]{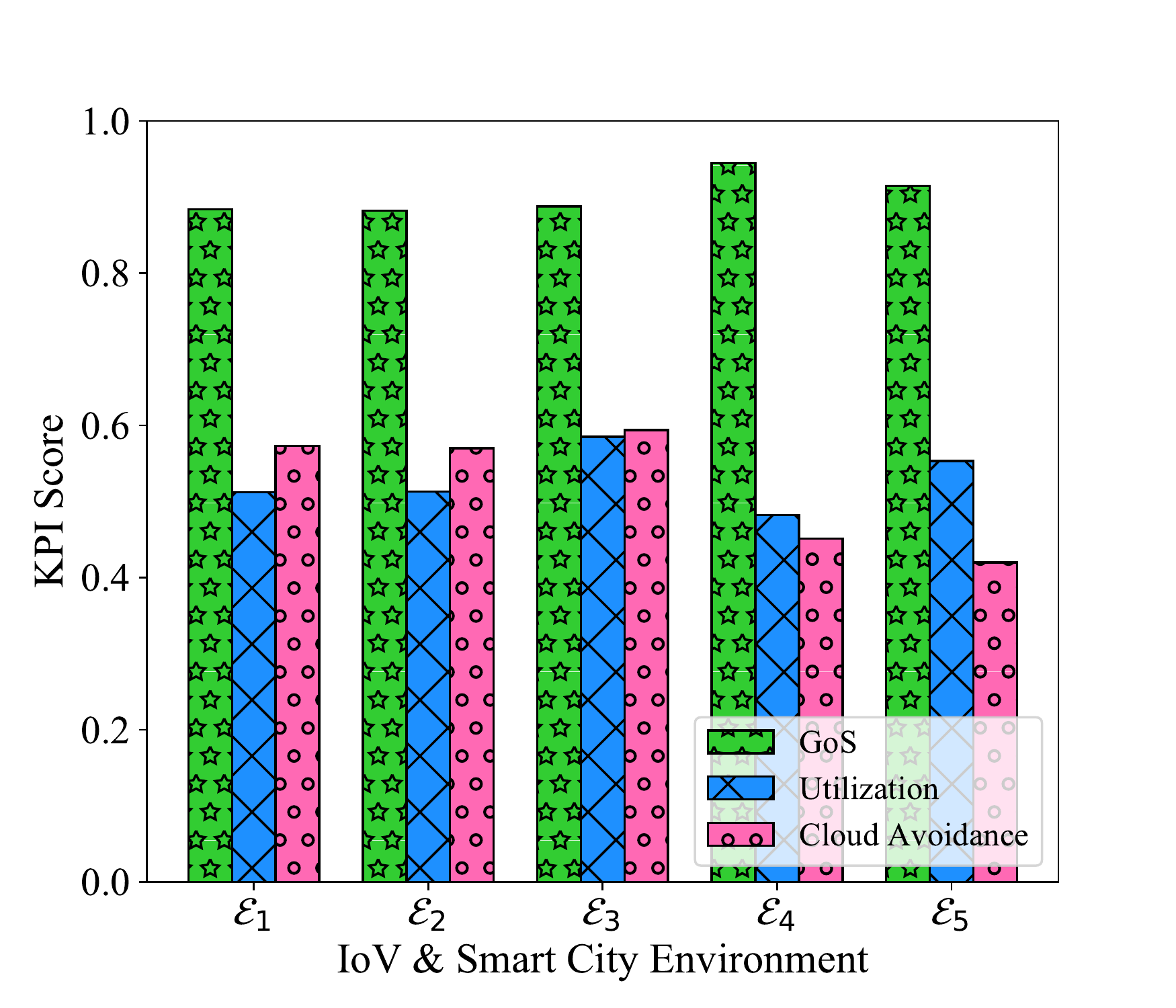}
		\caption{DQN-EC, $\omega_g=0.7$ and $\omega_u=0.3$.}
		\label{f:dqn70}
	\end{subfigure}
	\begin{subfigure}{.3\textwidth}
		\includegraphics[width=\linewidth]{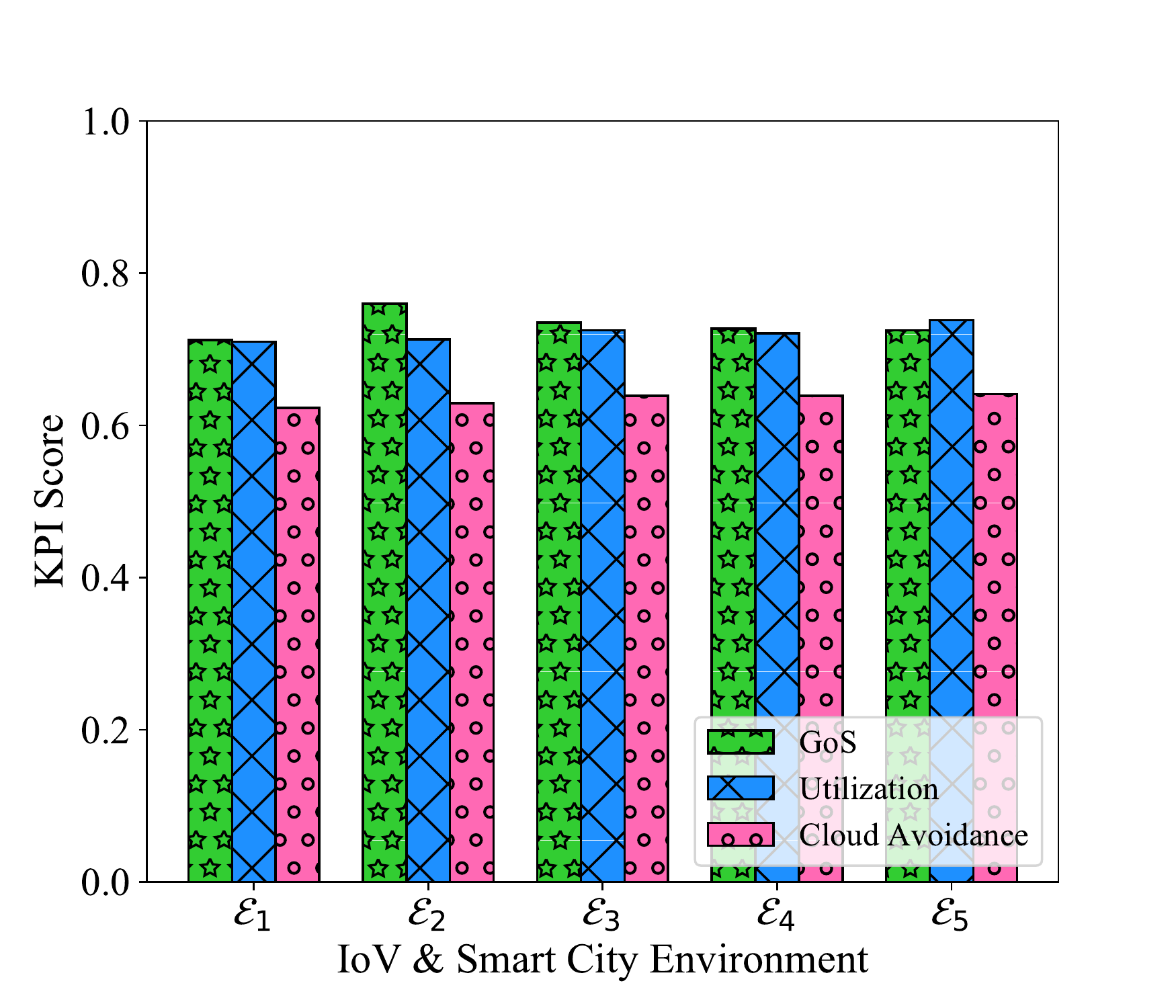}
		\caption{DQN-EC, $\omega_g=\omega_u=0.5$.}
		\label{f:dqn50}
	\end{subfigure}
	\begin{subfigure}{.3\textwidth}
		\includegraphics[width=\linewidth]{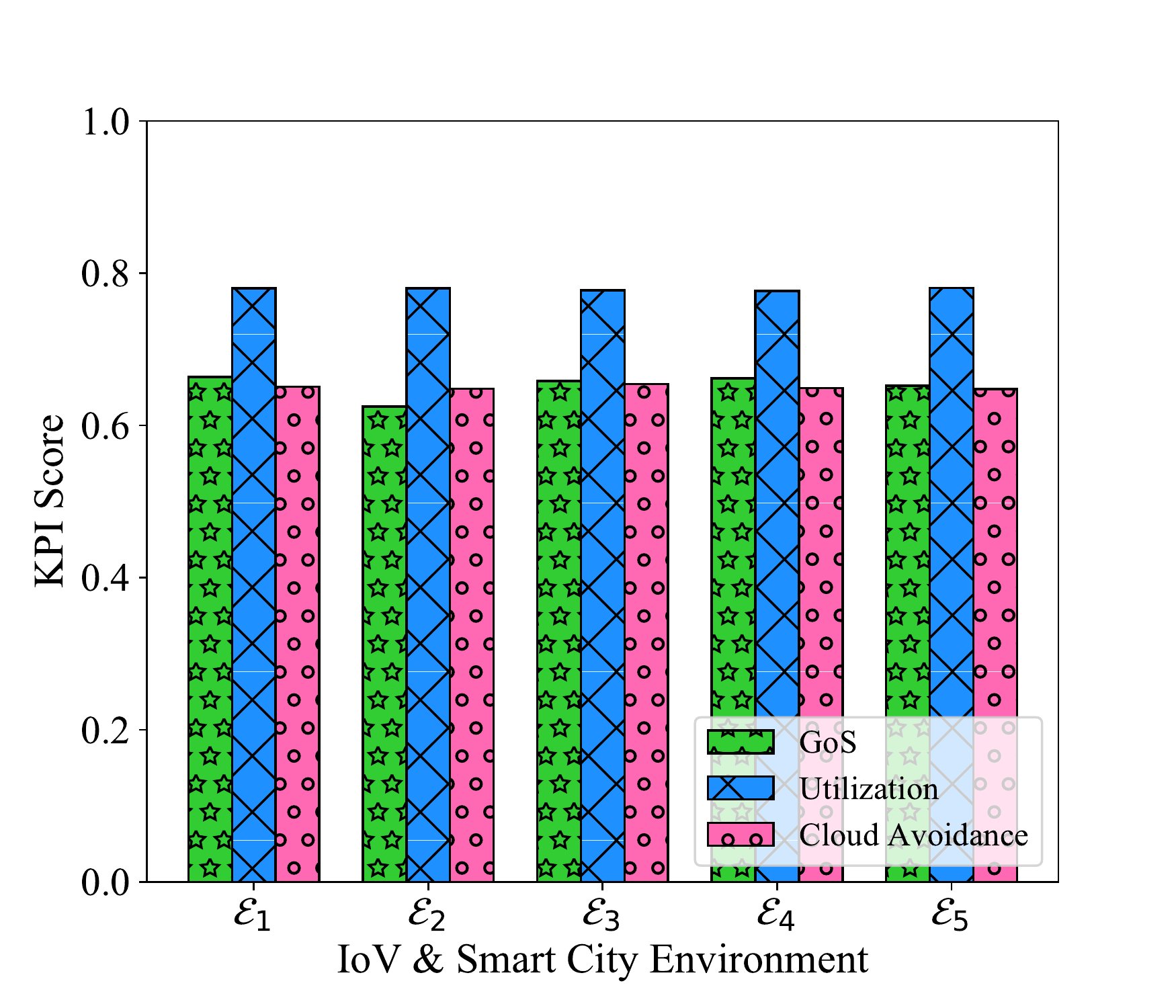}
		\caption{DQN-EC, $\omega_g=0.3$ and $\omega_u=0.7$.}
		\label{f:dqn30}
	\end{subfigure}\\
	\begin{subfigure}{.3\textwidth}
  		\includegraphics[width=\linewidth]{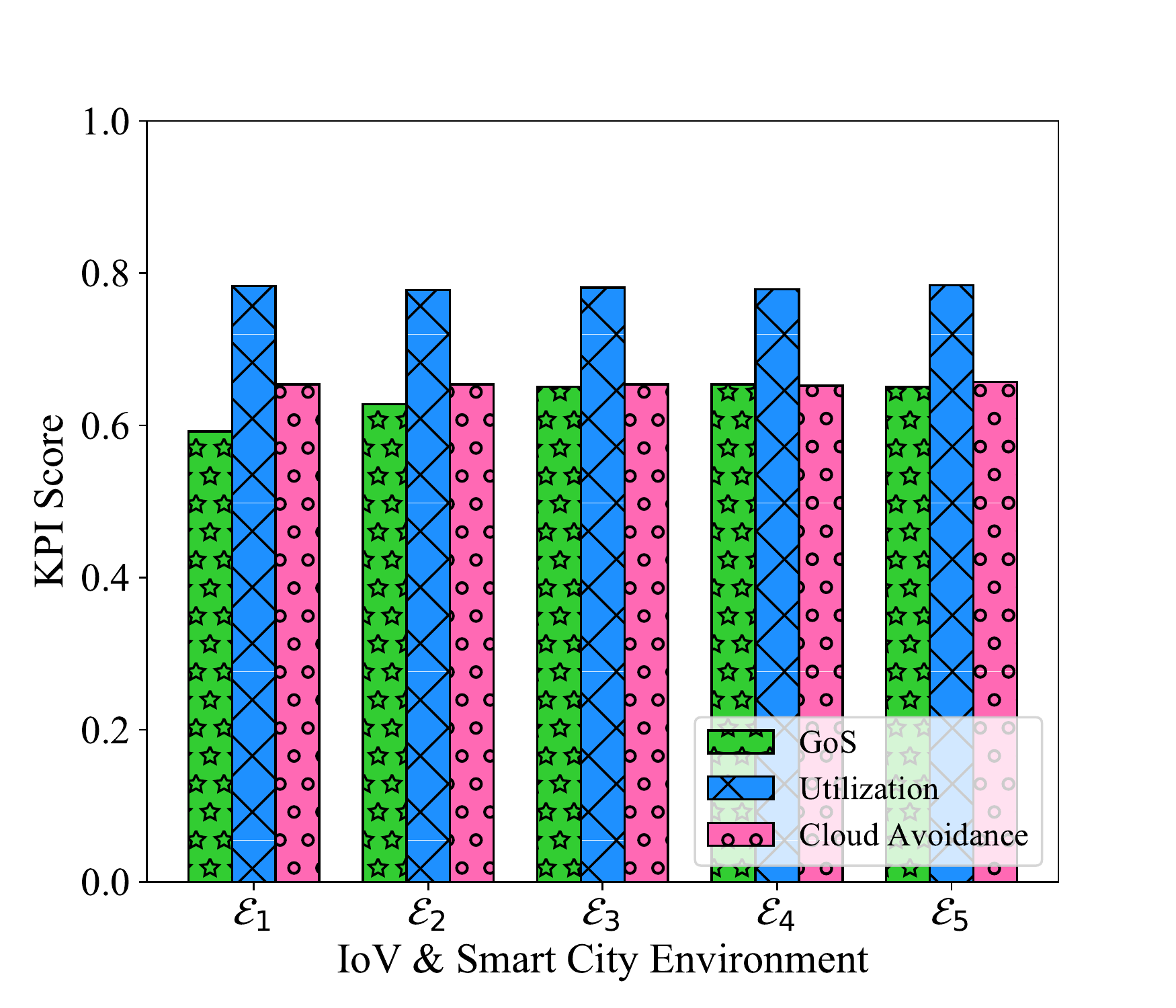}
		\caption{SAU-EC, all scenarios.}
		\label{f:sau}
	\end{subfigure} \hspace{5mm}
	\begin{subfigure}{.3\textwidth}
		\includegraphics[width=\linewidth]{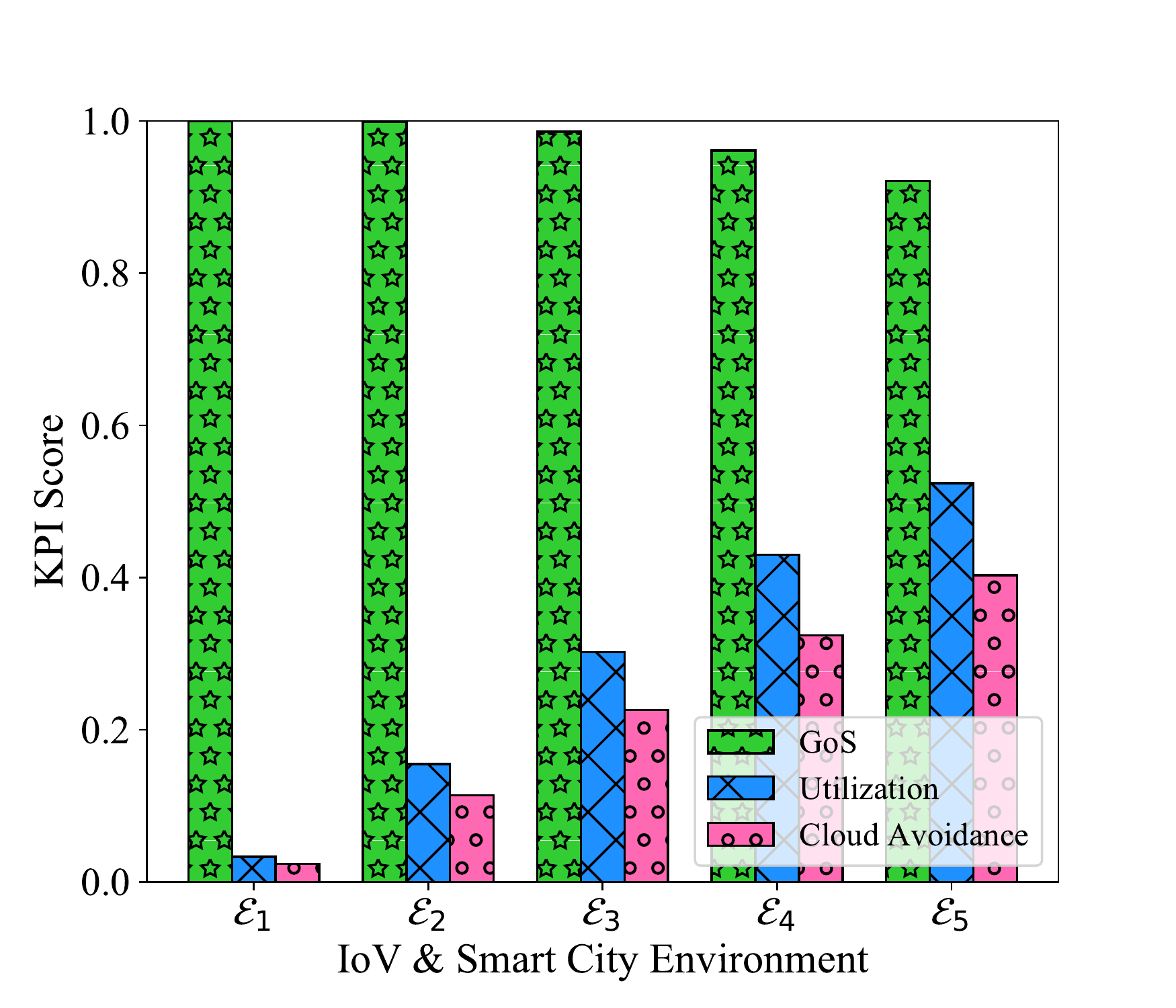}
		\caption{SHU-EC, all scenarios.}
		\label{f:shu}
	\end{subfigure}\\
	\begin{subfigure}{.3\textwidth}
		\includegraphics[width=\linewidth]{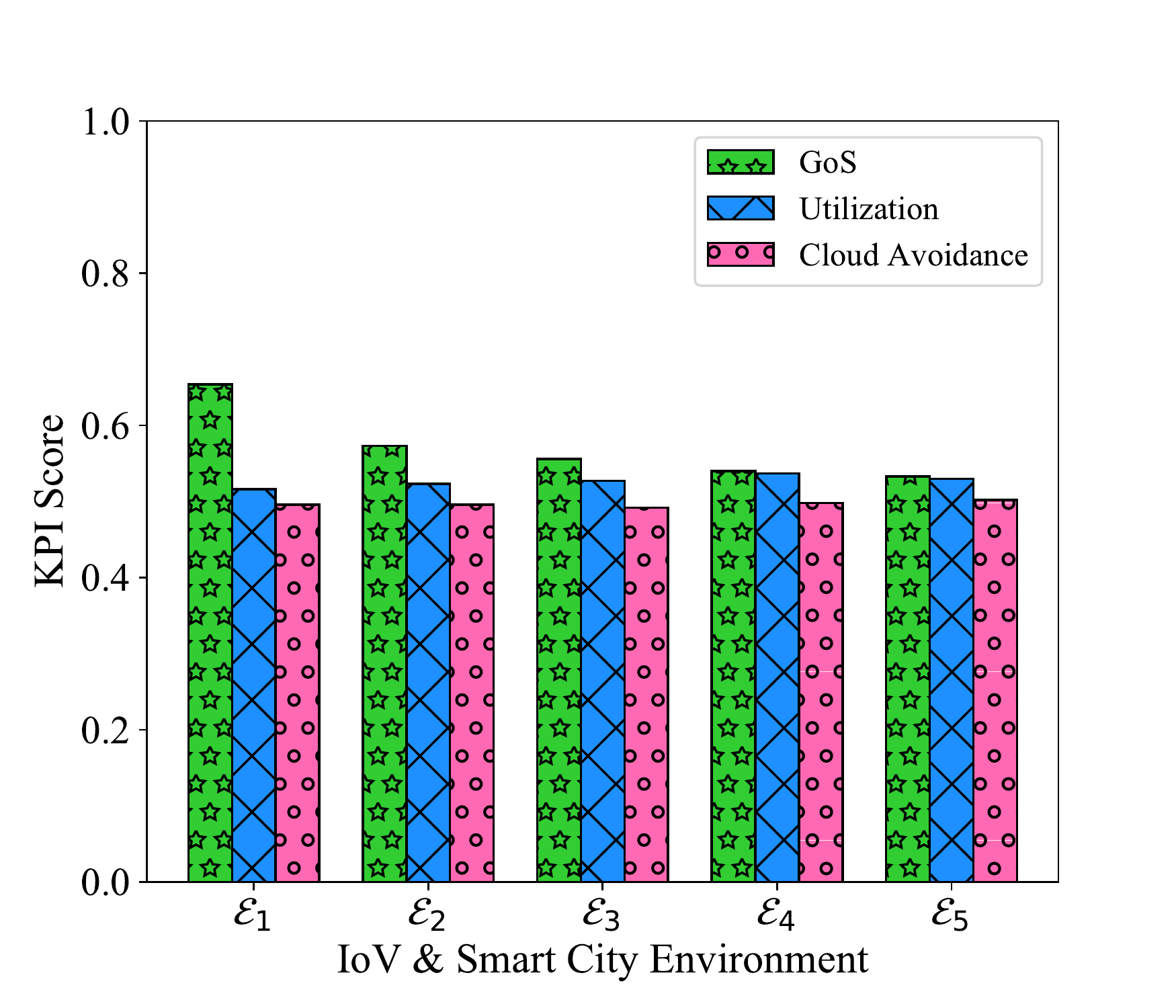}
		\caption{QL-NEC, $\omega_g=0.7$ and $\omega_u=0.3$.}
		\label{f:ql70}
	\end{subfigure}
	\begin{subfigure}{.3\textwidth}
		\includegraphics[width=\linewidth]{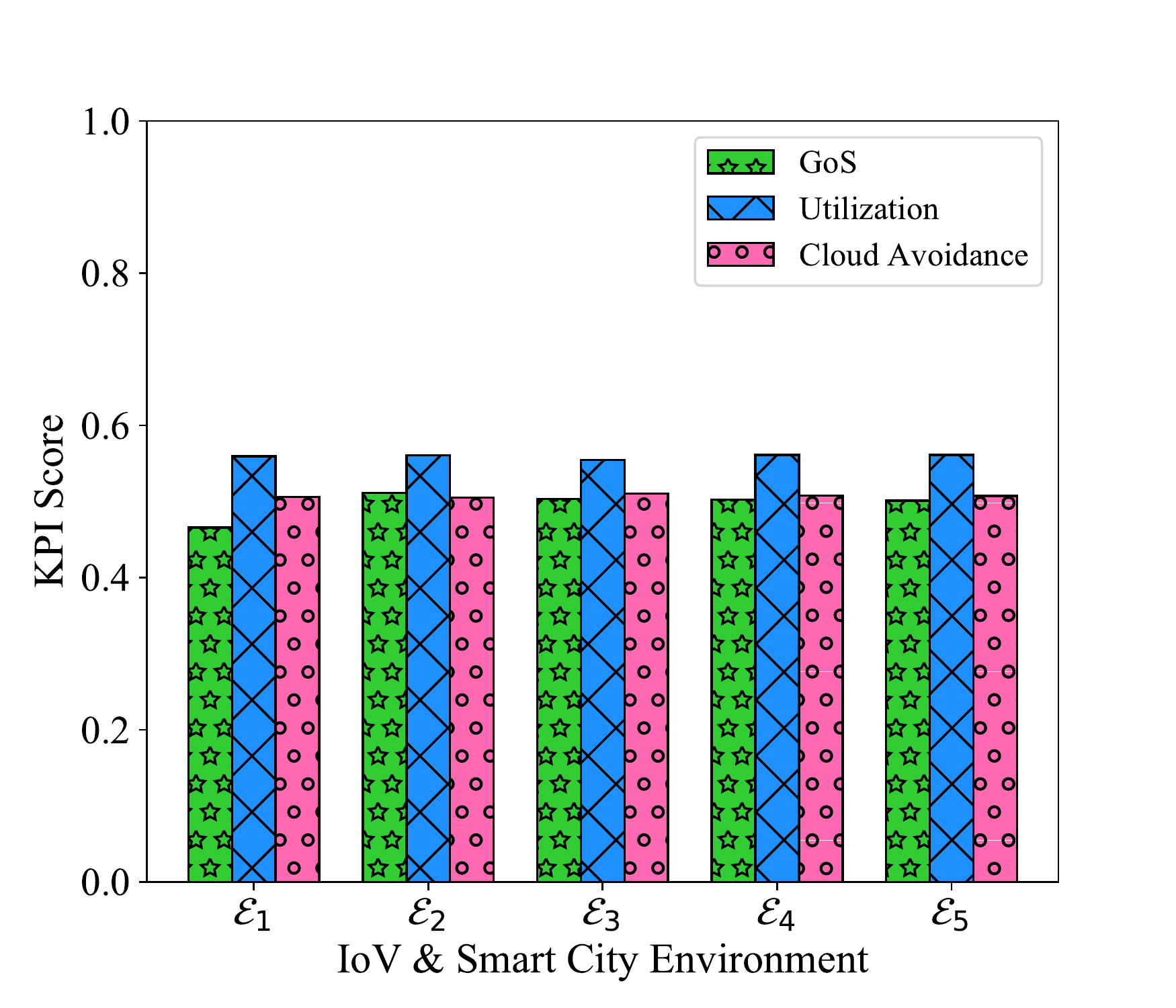}
		\caption{QL-NEC, $\omega_g=\omega_u=0.5$.}
		\label{f:ql50}
	\end{subfigure}
	\begin{subfigure}{.3\textwidth}
		\includegraphics[width=\linewidth]{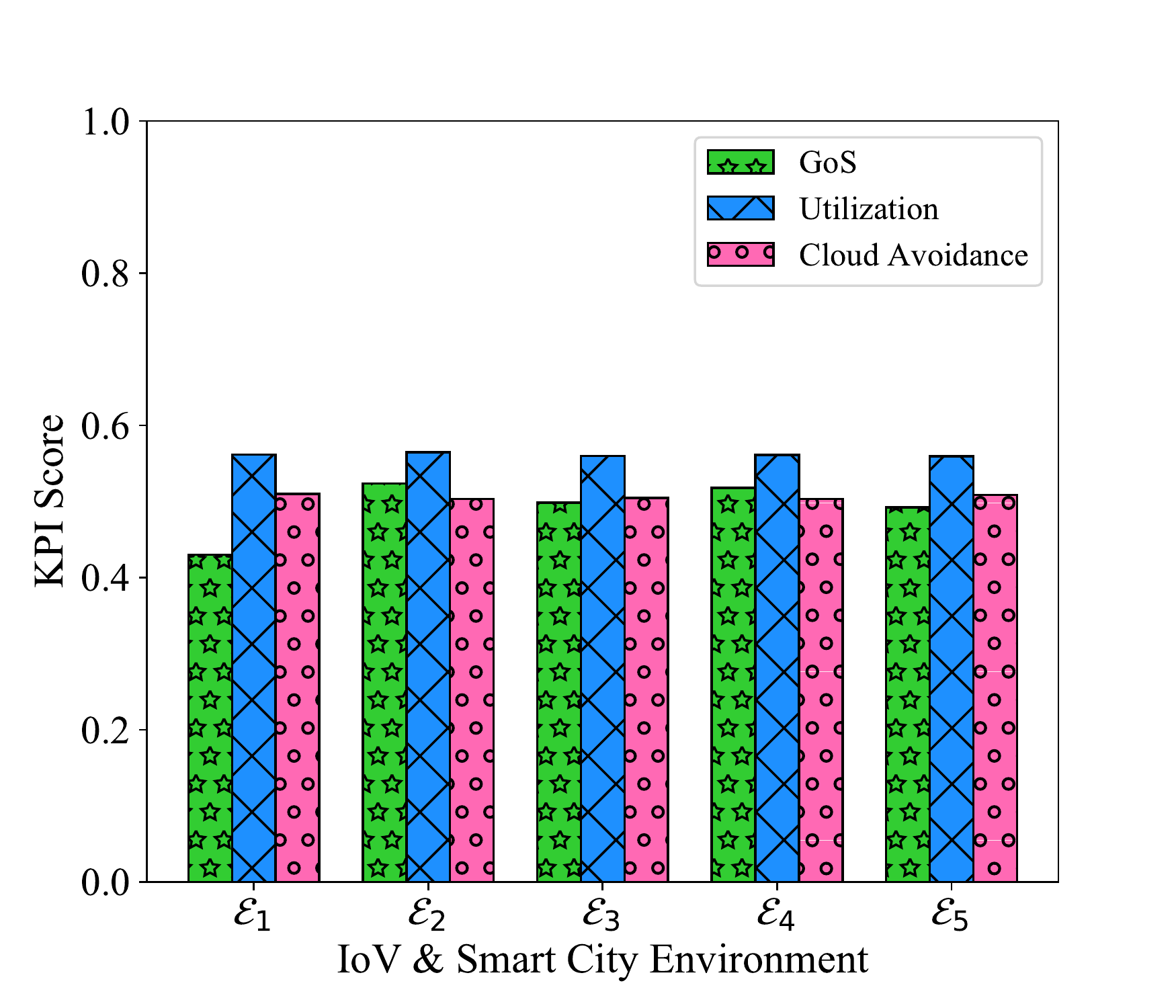}
		\caption{QL-NEC, $\omega_g=0.3$ and $\omega_u=0.7$.}
		\label{f:ql30}
	\end{subfigure}
	\caption{The score of the main three individual KPIs, GoS, resource utilization, and cloud avoidance when the EC applies Algorithm \ref{a:dqn}, SAU and SHU for coordinated FNs, and the uncoordinated QL at FNs with no EC, under the 3 scenarios in Table \ref{t:r}.}
	\label{f:kpi}
\end{figure*}

\subsection{Simulation Results}
\label{results}
We firstly consider the environment $\cE_3$ shown in Table \ref{t:env} and the scenario 1 in Table \ref{t:r}. By interaction with the environment as illustrated in Fig. \ref{f:dqn}, the EC learns the optimal policy using the DQN method given in Algorithm \ref{a:dqn}. Fig. \ref{f:convergence} shows the learning curve for the proposed DQN-based EC in terms of the overall performance and KPIs which quickly converge to the optimal scores. Starting with random actions for $30k$ time steps, the EC initially puts more weight on utilizing its resources and hence many high-utility requests are missed. However, as exploration rate decays, the EC quickly aligns with the objectives of scenario 1 putting more emphasis on GoS by prioritizing high-utility users for edge service. The converged score of cloud-avoidance KPI shows that edge slice serves the $50\%$ of the total received tasks.

Next, we compare DQN-EC given in Algorithm \ref{a:dqn} with SAU-EC, SHU-EC, and QL with no EC (QL-NEC) under the three scenarios given in Table \ref{t:r}. Figs. \ref{f:perf70}-\ref{f:perf30} show that the DRL-based EC adapts to each scenario and outperforms the other algorithms in all IoV and smart city environments. For scenario 1 in Fig. \ref{f:perf70}, SHU-EC has a comparable performance to DQN-EC because SHU algorithm promotes serving high-utility requests all the time, which matches with the focus on GoS in scenario 1 design objective with $\omega_g=0.7$. However, in poor IoV and smart city environments with less high-utility population such as $\cE_1$ the performance gap increases. This gap shrinks as environment becomes richer and SHU-EC achieves a performance as high as the DQN-EC score in $\cE_4$ and $\cE_5$. The performance of SAU-EC slightly increases while moving from $\cE_1$ to $\cE_3$ and becomes stable afterwards even for the richer environments $\cE_4$ and $\cE_5$ since SAU-EC does not prioritize high-utility tasks. Unlike the other algorithms, QL-NEC shows a declining trend since the network slicing problem becomes more challenging with uncoordinated FNs while moving towards richer environments in this scenario. Fig. \ref{f:perf50} represents scenario 2 with equal weights for GoS and resource utilization, where SAU-EC is the second performing algorithm following DQN-EC. With less importance for GoS, the performance of SHU-EC is as low as the QL-NEC in $\cE_1$ and although it grows while moving to richer environments, it does not reach a comparable level until $\cE_4$ and $\cE_5$. The uncoordinated FNs with QL-NEC is more steady in scenario 2. Fig. \ref{f:perf30} shows the performances in scenario 3 in which more emphasis is put on resource utilization than GoS with $\omega_u=0.7$. It is observed that SHU-EC fails to achieve a comparable level of performance compared to DQN-EC while SAU-EC does.

Fig. \ref{f:kpi} provides the detailed KPI scores for GoS, resource utilization and cloud avoidance for all algorithms considering the three design scenarios in all environments. DQN-EC always adapts to the design objective and the IoV and smart city environment. It maximizes GoS in scenario 1 as shown in Fig. \ref{f:dqn70}, balances GoS and utilization for scenario 2 as observed in Fig. \ref{f:dqn50}, and promotes resource utilization for scenario 3 as shown in Fig. \ref{f:dqn30}. QL-NEC in Figs. \ref{f:ql70}-\ref{f:ql30} tries to behave similarly as it learns by interaction, but unfortunately the uncoordinated FNs in the edge slice cannot achieve that. Note that, DQN-EC learns the right balance between GoS and resource utilization in each scenario. For instance, even though SHU-EC is the second performing in Fig. \ref{f:perf70} following DQN-EC, it has lower utilization and cloud avoidance scores, i.e., less edge-slice contribution to handle service requests as shown in Fig. \ref{f:shu}. Similarly, SAU-EC is well-performing in scenario 2 compared to DQN-EC as shown in Fig. \ref{f:perf50}, however, it does not learn to balance GoS and utilization as DQN-EC does in Fig. \ref{f:dqn50}.

\begin{figure}[!]
\centering
\includegraphics[width=.45\textwidth]{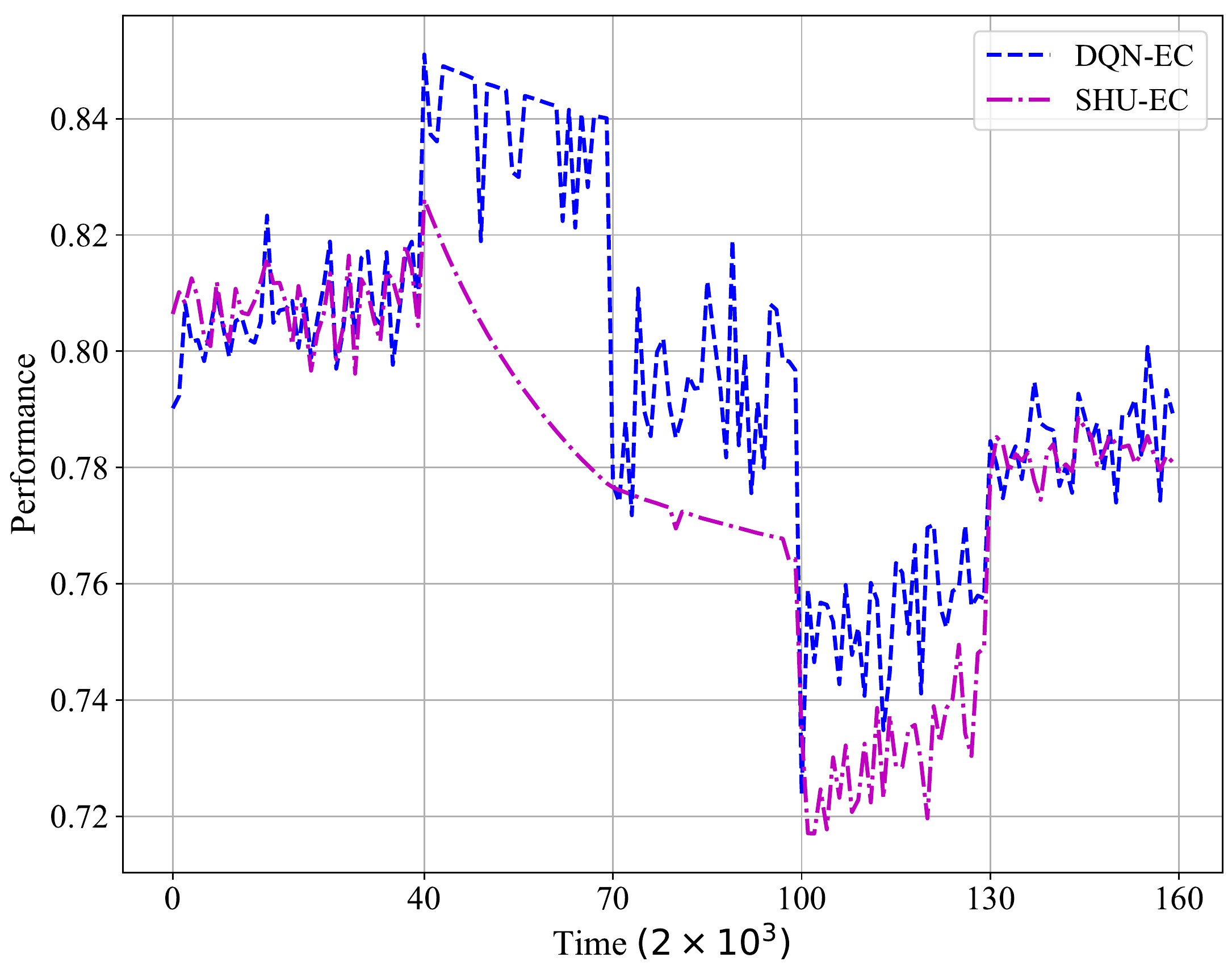}
\caption{The performance of the proposed DQN and the straightforward SHU policy for network slicing in a dynamic IoV and smart city environment considering the design objective of scenario 1 in Table \ref{t:r}. Although SHU performs well in rich environments, it cannot adapt to the other environments as expected. The proposed DQN policy on the other hand learns to adapt to different environments.}
\label{f:adaptivity}
\end{figure}

Finally, we test the performance of the proposed DQN algorithm in a dynamic IoV and smart city environment. In Fig. \ref{f:adaptivity}, we consider the design objectives of scenario 1 in Table \ref{t:r} and a sampling rate of $5\times10^{-4}$. To generate a dynamic IoT environment we start with $40$ samples for the initial environment and then change $\cE$ every $30$ samples. More samples is considered for the initial $\cE$ since we start with vacant resource blocks for all FNs in the edge slice.
We consider a dynamic IoV and smart city environment whose composition of high-utility requests, i.e., low-latency tasks, changes over a day. Starting in the morning busy hours with $\cE_4$, the density of high-utility requests drops over time to $\cE_1$ at the late morning hours after which it starts growing to reach $\cE_2$ by noon and $\cE_3$ in the evening, and then peaks again towards night busy hours with $\cE_5$. These 5 environments represent different distributions for a diverse levels of utilities, i.e., different latency requirements of the various IoV and smart city aforementioned applications, hence they can be thought as different traffic profiles and busy hours in terms of the required QoS in IoV and smart city. These busy hours directly affect the overall distribution of  the environment over time and makes it dynamic. In the proposed algorithm, once the EC detects the traffic profile, i.e., the environment, it applies the corresponding optimal policy $\pi_4^*$ given in \eqref{e:aa^*} to maximize the expected rewards in $\cE_4$. Right after the density of low-latency tasks drops over time to $\cE_1$, i.e., at $t=80k$, the EC keeps following $\pi_4^*$ until it detects the change from the statistics of task utilities, which results in a slight degradation in its performance since $\pi_4^*$ is no longer optimal for the new environment $\cE_1$. However, after a short learning period, the EC adapts to the dynamics, and switches to the new optimal policy $\pi_1^*$. Similarly, as seen for the other transitions from $\cE_1$ to $\cE_2$, $\cE_2$ to $\cE_3$, and $\cE_3$ to $\cE_5$, DQN-EC successfully adapts to the changing IoV and smart city environments. Whereas, the straightforward SHU-EC policy performs well in only the rich environments for which it was designed, and cannot adapt to the changes in the environment as expected. 

\section{Conclusion}
\label{conclusion}
We developed an infinite-horizon Markov decision process (MDP) formulation for the network slicing problem in a realistic fog-RAN with cooperative fog nodes; and proposed a deep reinforcement learning (DRL) solution for the edge controllers (ECs), which are the fog nodes that serve as cluster heads, to learn the optimal policy of allocating the limited edge computing and processing resources to vehicular and smart city applications with heterogeneous latency needs and various task loads. The deep Q-Network (DQN) based EC quickly learns the dynamics through interaction with the environment and adapts to it. DQN-EC dominates the straightforward and non-cooperative RL approaches as it always learns the right balance between GoS and resource utilization under different performance objectives and environments. In a dynamic environment with changing distributions, DQN-EC adapts to the dynamics and updates its optimal policy to maximize the performance.

\FloatBarrier
\bibliographystyle{IEEEtran}
\bibliography{refs}
\end{document}